\documentclass[draft,usenatbib]{mn2e}
\usepackage{psfig,amsmath}

\def\gsim{\mathrel{\raise0.35ex\hbox{$\scriptstyle >$}\kern-0.6em 
\lower0.40ex\hbox{{$\scriptstyle \sim$}}}}
\def\lsim{\mathrel{\raise0.35ex\hbox{$\scriptstyle <$}\kern-0.6em 
\lower0.40ex\hbox{{$\scriptstyle \sim$}}}}
\def\oii{{\rm O{\sc ii}}}

\def\Mpc{\,\hbox{Mpc}}

\def\arcsec{\hbox{$^{\prime\prime}$}}
\def\arcmin{\hbox{$^{\prime}$}}

\def\fp{\hbox{f}_{\rm p}}
\def\fhsf{\hbox{f}_{\rm hsf}}

\def\Mstar{\hbox{M}_{*}}
\def\Bj{$B_{\rm J}$}

\def\MBj{$M_{B_{\rm J}}$}
\def\Halpha{$\rm H\alpha$}
\def\ewoiim{\raise0.35ex\hbox{$\scriptstyle <$}\hbox{\rm EW[OII]}\raise0.35ex\hbox{$\scriptstyle >$}}
\def\ewoiimsf{\raise0.35ex\hbox{$\scriptstyle <$}\hbox{\rm EW[OII]}{\LARGE\hbox{$\scriptstyle |$}}\hbox{SF}\raise0.35ex\hbox{$\scriptstyle >$}}
\def\kmsMpc{$\rm kms^{-1}Mpc^{-1}$}

\def\BC{\hbox{R}_{\rm (B/C)}}
\def\BF{\hbox{R}_{\rm (B/F)}}

\def\dBC{\hbox{$\Delta$R}_{\rm B/C}}
\def\dBF{\hbox{$\Delta$R}_{\rm B/F}}
\def\SBC{\hbox{S}_{\rm B/C}}
\def\SBF{\hbox{S}_{\rm B/F}}
\def\svo {\sigma(v)_{obs}}
\def\svr {\sigma(v)_{rest}}
\def\svi {\sigma(v)_{intr}}
\def\dv {\Delta(v)}
\def\dz {\delta(z)}
\def\dth {\delta(\theta)}
\def\dr {\delta(r)}

\begin{document}

\title{Galaxy Groups at ${\bf 0.3 \leq z \leq 0.55}$. I. Group Properties}

\author[D.~J.~Wilman et al.]
{D.~J.~Wilman$^{1,2,6}$,~M.~L.~Balogh$^{1,3}$,~R.~G.~Bower$^1$,~J.~S.~Mulchaey$^4$,
\and
  ~A.~Oemler~Jr$^4$,~R.~G.~Carlberg$^5$,~S.~L.~Morris$^1$,~R.~J.~Whitaker$^1$\\
$^1$Physics Department, University of Durham, South Road, Durham DH1 3LE, U.K.\\
$^2$Max-Planck-Institut f\"ur extraterrestrische Physik, Giessenbachstra\ss e, D-85748 Garching, Germany (present address)\\
$^3$Department of Physics, University of Waterloo, Waterloo, Ontario,
Canada N2L 3G1 (present address).\\
$^4$Observatories of the Carnegie Institution, 813 Santa Barbara Street, Pasadena, California, U.S.A.\\
$^5$Department of Astronomy, University of Toronto, Toronto, ON, M5S 3H8 Canada.\\
$^6$\emph{email: dwilman@mpe.mpg.de}}

\maketitle

\begin{abstract}
The evolution of galaxies in groups may have important implications for the
evolution of the star formation history of the universe, since many
processes which operate in groups may suppress star formation and the
fraction of galaxies in bound groups grows rapidly between $z=1$ and
the present day. In this paper, we present an investigation of the
properties of galaxies in galaxy groups at intermediate redshift
($z\sim0.4$). The groups were selected from the CNOC2 redshift survey as
described in \citet{Carlberg01}, with further spectroscopic
follow-up undertaken at the Magellan telescope in order to improve the
completeness and depth of the sample. We present the data for the
individual groups, and find no clear trend in the fraction of
passive galaxies with group velocity dispersion and group
concentration.  We stack the galaxy groups in order to compare the
properties of group galaxies with those of field galaxies at the same
redshift. The groups contain a larger fraction of passive galaxies
than the field, this trend being particularly clear for galaxies brighter
than $M_{B_J}<-20$ in the higher velocity dispersion groups.  
In addition, we see evidence for an excess of bright passive galaxies
in the groups relative to the field. In contrast, the luminosity functions
of the star forming galaxies in the groups and the field are consistent. 
These trends are qualitatively consitent with the differences between group
and field galaxies seen in the local universe.
\end{abstract}


\section{Introduction}

Clusters of galaxies have received intensive observational study over
the last decades.  This effort has lead to clear 
results on the way galaxy properties, such as morphology and
star formation rate, vary within clusters and how these properties evolve 
between clusters at different redshifts \citep[][see Bower \& Balogh
2004\nocite{BowerReview04} for a recent
review]{Dres80disc,BO84,Dressler97,Poggianti99}. In contrast, comparably--detailed studies
of galaxy groups and  
their evolution have only recently begun in earnest.
The group environment is likely to have a significant
impact on star formation rates in the member galaxies \citep{Zabludoff98,Hashimoto98,Tran01}
and recent work has emphasied that even for galaxies now in rich 
clusters, much of the transformation of the galaxies' properties may have taken 
place in groups embedded in the filamentary structure
\citep{Kodama01,Balogh03}.
In addition, while few galaxies in the local universe are located in
clusters, up to 50 per cent \citep{HuchraGeller82,Eke04} may be located in
galaxy groups.  Furthermore, this number is strongly
redshift--dependent, as   a larger and larger fraction of galaxies
become members of groups as the large-scale
structure of the universe develops.
Thus the properties of galaxies in groups and the impact
of the group environment on the evolution of galaxies can have an important
bearing on the decline of the cosmic star formation rate from $z=1$
to the present-day \citep{Lilly96,Madau98,Hopkins04}.

Studies of nearby groups show that their galaxy populations exhibit properties which vary from cluster-like
(mostly early-type) to field-like (mostly late type) \citep{Zabludoff98}. 
However we know that groups span
a wide range in local density, upon which the morphological composition 
\citep{PostmanGeller84,Zabludoff98,Tran01} and the
mean star forming properties \citep{Hashimoto98,Balogh03} strongly depend
\citep[e.g.][]{Dres80disc,PostmanGeller84,Hashimoto98,Gomez03,Balogh03,Kauffmann04}. 
The wide range of galaxy populations found in nearby groups is likely to be a natural 
consequence of these correlations. 
Conversely, the powerful dependence of galaxy properties on local densities typical of groups suggests that
groups provide the ideal environment for galaxy transformations. This may result in 
the strong dependence of early type fraction and passive dwarf galaxy abundance on group velocity dispersion
and X-ray luminosity, observed in nearby groups \citep{Zabludoff98,Zabludoff2000,Christlein2000}.
In particular, galaxy interactions are expected to be 
common in groups, in which the velocity dispersion
is typically not much larger than that of the constituent galaxies \citep{Hashimoto98,SS01}. 

Groups have a much lower density contrast against the background
galaxy population, so
most work has concentrated on groups which
are either unusually compact \citep{Hickson89} 
or X-ray luminous \citep{Mulchaey03}. More recently, the
advent of large field galaxy redshift surveys has made it
possible to study galaxy groups selected purely on the basis of their
three dimensional galaxy density.  In the local universe, early redshift surveys and 
more recently the extensive
2dF Galaxy Redshift Survey (2dfGRS) and Sloan Digital Sky Survey (SDSS) have generated large group catalogues
which can be used as the basis for studies of galaxies in the 
group environment 
\citep{HuchraGeller82,Ramella89,Ramella97,Ramella99,Hashimoto98,Tucker2000,Martinez02,Eke04,Balogh03}
However, 
the properties of higher redshift groups have been
relatively little explored.  This is largely because of the difficulty
in finding suitable systems, and the low success rate of spectroscopic
follow-up of group members. \citet{AllSmith93} used radio
galaxies to preselect groups in order to study the evolution of the
blue galaxy fraction with redshift, while \citet{Carlberg01} have 
presented a group catalogue based on the CNOC2 galaxy redshift
survey.

In this paper, we present new data obtained in the region of 26
CNOC2 groups at $0.3 \leq z \leq 0.55$ with LDSS2 on Magellan. In
Sections~\ref{sec:CNOC2GroupData} and~\ref{sec:whyOIIforSF} we
introduce the data and explain why we have chosen to use the [OII]
emission line to study statistical trends in star formation. In
Section~\ref{sec:CNOC2DataProcessing} we begin by describing the data
reduction procedure and our method for determination of redshifts and
emission line equivalent widths for each
galaxy. Section~\ref{sec:grpmem} then goes on to explain our procedure
for group membership allocation. In Section~\ref{sec:Indivgrps} we
present the properties of the 26 individual groups in order to examine
any trends of structural group properties with star formation. Finally,
in Section~\ref{sec:stackedgrp} groups are stacked to enable detailed
statistical analaysis, and we investigate the link between the group
environment and star formation by contrasting the stacked group with
our sample of field galaxies selected from the same redshift
range.

In a second paper \citep[][hereafter known as 
  Paper~II]{Wilman04b}, we present a comparison of the star forming
properties of these intermediate redshift groups with local groups
selected from the 2PIGG catalogue at $0.05 \leq z \leq 0.1$
\citep{Eke04}.

Throughout this paper we assume a $\Lambda$CDM cosmology of $\Omega_{M}
= 0.3$, $\Omega_{\Lambda}=0.7$ and $H_\circ = 75$\kmsMpc.  

\section{CNOC2 groups}\label{sec:CNOC2GroupData}

\subsection{The survey}\label{sec:CNOC2survey}

The second Canadian Network for Observational Cosmology Redshift Survey
(CNOC2) was recently completed with the aim of studying galaxy
clustering in the redshift range $0.1<z<0.6$
\citep{Yee00,Shepherd01}. The CNOC2 survey is split into 4 patches,
approximately equally spaced in RA and totalling 1.5 square degrees in
area. The survey consists of 5 colour $UBVR_{C}I_{C}$ photometry with
$\sim  4\times10^{4}$ galaxies down to the photometric limit $R_{C}=
23.0$. The MOS spectrograph on the CFHT 3.6m telescope was used to
obtain spectra for over $\sim6000$ galaxies in total, 48 per cent complete
down to $R_{C}=21.5$. Combination of the imaging and spectroscopy lead
to a very well determined selection function for the spectroscopic
sample \citep{Yee00} and comparisons with our deeper spectroscopy
suggest that brighter than this limit the CNOC2 survey is not biased
towards emission line objects. The transmission efficiency of the band
limiting filter and grism combination was above half power in the
wavelength range $4387-6285$\AA, effectively limiting the redshift
range of the survey. The most prominent absorption features (Ca II H
and K) lie in this wavelength range for galaxies in the redshift range
$0.12<z<0.58$, whereas either the [OII] or [OIII] emission lines fall
in this wavelength range for all galaxies in the range $0<z<0.68$.  

\subsection{The CNOC2 groups}\label{sec:CNOC2groupsIntro}

Distant groups have always been difficult to recognize because of their
sparse galaxy populations. The presence of high redshift groups has
typically been inferred indirectly via the presence of a radio galaxy
or X-ray emitting intragroup medium 
\citep[IGM,][]{AllSmith93,Jones02b}. Whilst these surveys provide a useful
insight into the evolution of galaxy properties in rich galaxy groups,
the selection criteria strongly bias the selection towards these
richer elliptical dominated groups, whereas low redshift samples point
to a more numerous population of low density, X-ray faint, spiral
dominated groups \citep{Mulchaey03}.

The CNOC2 survey provided an powerful opportunity to generate a
kinematically selected sample of galaxy groups. A
friends-of-friends percolation algorithm was used to detect
groups of galaxies in redshift space. This was followed by a trimming
step in which a centre, velocity dispersion, $\sigma_{1}$ and $r_{200}$
($\sim$ virial radius) were estimated and then members were added or
deleted within $3\times \sigma_{1}$ and $1.5\times r_{200}$ in 3 rounds
of iteration \citep[see][]{Carlberg01}. A total sample of over 200
galaxy groups was detected. Although the spectroscopic sample is
incomplete, this group sample represents a kinematically selected
sample at intermediate redshifts ($0.12<z<0.55$), free from the strong
biases present in samples selected by other means.  

\subsection{Deeper LDSS2 data}\label{sec:LDSS2CNOC2data}

We used the Multi-Object Spectrograph LDSS2 (Low Dispersion Survey
Spectrograph) on the 6.5m Baade Telescope at Las Companas Observatory
(LCO) in Chile to obtain a deeper and more complete sample of galaxies
in the region of 20 of the CNOC2 groups at $0.3 \leq z \leq 0.55$
located in the 3 out of 4 CNOC2 patches accessible from the latitude of
LCO. Masks were designed using an automated selection algorithm to
minimise selection bias and maximise the allocation of targets per
mask. Due to the proximity on the sky of some of the CNOC2 groups, the
masks were chosen to serendipitously sample a further 11 groups from
the \citet{Carlberg01} sample, although only 6 of these lie at $z >
0.3$. Targets were prioritised using an automated algorithm which
favours brighter objects and objects which lie close together along the
mask's spectral axis. This second criterion ensures that the spectral
coverage does not vary wildly from object to object in each mask. The
data were taken in 4 separate observing runs between May 2001 and
November 2002. Each LDSS2 mask covers approximately 6.5\arcmin\
$\times$ 5\arcmin\ and in the CNOC2 fields it is normal to fit between
20 and 30 slits onto a mask. Between 1 and 3 masks were observed per
target group, depending upon the density of target galaxies in that
area of sky. Mask exposures varied from 1 to 4 hours depending on the
phase of the moon and observing conditions. For maximum throughput, the
blue-optimised medium resolution grism was used with a dispersion of
5.3\AA\ (on the fourth run the red optimised grism was used which has
the same dispersion and similar efficiency at wavelengths of
interest). The slits have a width $= 1.47$\arcsec which corresponds to
$8.78$\AA\ of spectral coverage (compared to $6.28$\AA\ in the original
CNOC2 masks). Using the CNOC2 photometric classifier from the CNOC2
photometric catalogue, we targeted 634 objects classified as
\emph{galaxies}, 102 objects classified as \emph{possible
  galaxies}. Where there were free spaces on the mask with no galaxies
present, we allocated 130 objects classified as \emph{probable
  stars}. From this last category, 29 out of 130 objects had
galaxy-like spectra which yielded redshifts. The remaining 101 were
correctly identified as stars. Included in our targets were 35 galaxies
which already had redshifts from the initial CNOC2 survey. These were
reobserved to form a comparison sample which is used to understand the
accuracy of our measurements.  

\begin{table}
\begin{center}
\caption{The number of galaxies targetted in each targetted group and the $0.3 \leq z \leq 0.55$ \citet{Carlberg01} groups serendipitous in these fields.}
\label{tab:obs}
\vspace{0.1cm}
\begin{tabular}{ccc}
\hline\hline
\noalign{\smallskip}
Targetted & Number of& Serendipitous \cr
Group & targetted & \citet{Carlberg01}\cr
& galaxies& groups\cr
\hline
22 & 24 & 23,33\cr
24 & 45 & 8\cr
25 & 48 & none\cr
28 & 86 & none\cr
31 & 38 & none\cr
32 & 20 & 13\cr
34 & 55 & none\cr
37 & 24 & 1,27,35\cr
38 & 48 & 40\cr
39 & 88 & 23,33\cr
134 & 47 & 132,138\cr
137 & 52 & 140\cr
138 & 54 & 129,132,133,134\cr
139 & 28 & none\cr
140 & 62 & 137\cr
227 & 53 & none\cr
228 & 19 & none\cr
232 & 24 & 241\cr
241 & 14 & 201,232\cr
244 & 37 & 243\cr
\noalign{\hrule}
\end{tabular}
\end{center}
\end{table}

In Table~\ref{tab:obs} we show the number of galaxies targetted in each
field and the serendipitous \citet{Carlberg01} groups which lie in
these fields. 

\section{Measurement of star formation using EW[\oii]}\label{sec:whyOIIforSF}

To study the relative levels of star formation in statistical galaxy
samples, we use the [OII]$\lambda3727$ emission line which lies
centrally in the visible window in our CNOC2 redshift range of $0.3\leq
z \leq 0.55$ and at wavelengths of low sky emission. At $z>0.21$, the
\Halpha\ emission line disappears entirely from the LDSS2 spectrograph
window with the instrument sensitivity dropping in the red with the
current optics and detector. Furthermore, at $z \gsim 0.1$ measurements of \Halpha\
are compromised by the increase in sky line density with increasing
wavelength. \citet{Hopkins03} have recently shown using data from local
galaxies in SDSS that measurements of SFR using [OII] emission are
consistent with those from \Halpha\ and 1.4GHz
  luminosities. Also the scatter in the \Halpha-[OII] relationship is
primarily luminosity dependent \citep{Jansen01} and so in a given
luminosity range the systematic error in using the [OII] measurement to
infer star formation rates is significantly reduced. Additionally, even
at $z<1$, \citet{Flores03} place the contribution from dusty starburst
galaxies to the total SFR, underestimated by optical emission line
measurements at $\lsim 20$ per cent. Rather than inferring star formation
rates, we deliberately limit our study to direct comparison between
[OII] measurements to avoid model dependency in our results. Finally,
we restrict our analysis to EW[OII] rather than [OII] flux and star
formation rates. Normalisation by the continuum reduces uncertainties
related to absorption by dust and aperture bias as well as providing a
measure of SFR per unit luminosity. 

\section{Data Processing}\label{sec:CNOC2DataProcessing}

\subsection{Data Reduction}\label{sec:CNOC2DataReduction}

We extracted and calibrated our spectra using mainly {\sc iraf} tools
in the \emph{onedspec} and \emph{twodspec} packages. Wavelength
calibration was applied using both {\sc iraf} tools and new MOS
reduction software written by Dan Kelson in Carnegie
\citep{Kelson03}. We have tested both these methods to ensure that
consistent solutions are obtained. The wavelength calibration is based
upon arcs taken during daytime and is secure in the wavelength range
$3700$\AA$\leq\lambda\leq8000$\AA\ which corresponds to $z\leq2.1$ for
the [OII] emission line and $z\leq0.6$ for [OIII] which extends far
beyond the redshift range of interest. Around bright skylines,
systematic residuals often remain. We interpolated over these regions. 

To account for mismatch between the daytime arc and nighttime science
observations, we then applied a zero point offset to the wavelength
calibration. This is computed by measuring the offset of the 5577\AA\
skyline in the arc-calibrated frames. Most of these offsets were $<
5$\AA\ and we did not detect any significant non-linearities. The new
wavelength solution appears robust, as evidenced in the comparison of
LDSS2 and CNOC2 redshifts (see Figure~\ref{fig:Zoiicomp} which is
explained in Section~\ref{sec:CNOC2cfLDSS2}). 

\subsection{Redshift Measurement}\label{sec:CNOCmeasurez}

The {\sc iraf} tool \emph{xcsao} in the \emph{rvsao} package
\citep{rvsao98} was used to cross-correlate all the spectra with
templates of rest-frame early-type (absorption line) and late-type
(emission line) galaxies. The early-type galaxy template was created by
coadding a large number of high signal to noise early type galaxy
spectra from the sample, each of which had been shifted to zero
redshift.\footnote{High signal to noise early type galaxy spectra used
  to create our template are shifted to zero redshift by
  cross-correlation with a high signal to noise spectrum of NGC3379
  from the atlas of \citet{Kennicutt92}.} The emission line (Sc)
template is a high signal to noise spectrum of NGC4775 from the atlas
of \citet{Kennicutt92} and smoothed to the resolution of
LDSS2.\footnote{Velocity zero points of both \citet{Kennicutt92}
  spectral templates have been improved to an accuracy of
  30~km~s$^{-1}$ \citep{Yee96}.} We shall henceforth refer to redshifts
obtained using the late-type template as \emph{emission} redshifts and
those obtained using the early-type template as \emph{absorption}
redshifts.

The peak in the cross-correlation spectrum is selected by \emph{xcsao}
to give an estimate of the redshift for each object \citep{TD79}. To
assess redshift measurements, redshifts attained using the early-type
(absorption) and late-type (emission) templates were assessed in each
case by manually inspecting the spectra and assigning a redshift
quality flag. The quality assessment is based upon the number of
believable emission lines in an emission line spectrum. In absorption
line spectra, we simply decide whether to accept or reject a
solution. For example, a solution with a visible 4000\AA\ break would
be accepted. Spectra containing a single emission line received further
checks to ensure the emission is real. Finally, redshifts which became
apparent upon visual inspection were applied by setting up that
solution as the initial guess in \emph{rvsao}. However, if no peak in
the cross-correlation spectrum could be obtained at the correct
redshift then the solution was abandoned.

The final redshift of a target was selected by comparing the redshift
quality flag from emission and absorption redshifts. Where the
absorption and emission redshifts agree within errors, emission
redshifts are preferred due to their greater accuracy. However, where
the emission redshift was not secure and a different, higher quality
absorption redshift exists, then the absorption redshift is
used. Objects with no acceptable redshift are discounted from the
spectroscopic sample. 

In total we have 418 redshifts from the LDSS2 data of which 240 are in
the range $21 \leq R_c \leq 22$ which dramatically improves the
completeness at this depth. In the whole sample, we have 86 new group
members from LDSS2 spectroscopy. This increases the total group sample
to 295 in 26 groups (see Group Membership section for further details
of group allocation). In the magnitude range $21 \leq R_c \leq 22$ and
within 150\arcsec\ of targetted group centres, we obtained 179
additional redshifts (36 members) on top of the 115 (27 members)
existing from the CNOC2 survey. We consider $R_{c}=22$ to be our
magnitude limit at which we can obtain redshifts for $\sim 60$ per cent of
objects targetted; our success rate for all targetted galaxies
brighter than $R_{c}=22$ 
is 74 per cent.
The selection functions are well understood and targetted
galaxies for which we do not manage to obtain a redshift are evenly
distributed in colour, suggesting that our sample is unbiased (see
Appendix~\ref{sec:completeness}).

\subsubsection{Emission Line Equivalent Width Measurements}\label{sec:CNOCMeasureEmLines}

We measure the [OII] equivalent widths using our own purpose written
code to compare the flux in two separate continuum regions with that in
a feature bandpass.  We use the definition from \citet{DressSchect87}
for [OII], in which the feature bandpass ranges from 3718\AA\ to
3738\AA\ and the continuua regions have width 65\AA\, adjacent to the line
region on either side.   

\subsubsection{Comparison with CNOC2 data}\label{sec:CNOC2cfLDSS2}

\begin{figure*}
\centerline{\psfig{figure=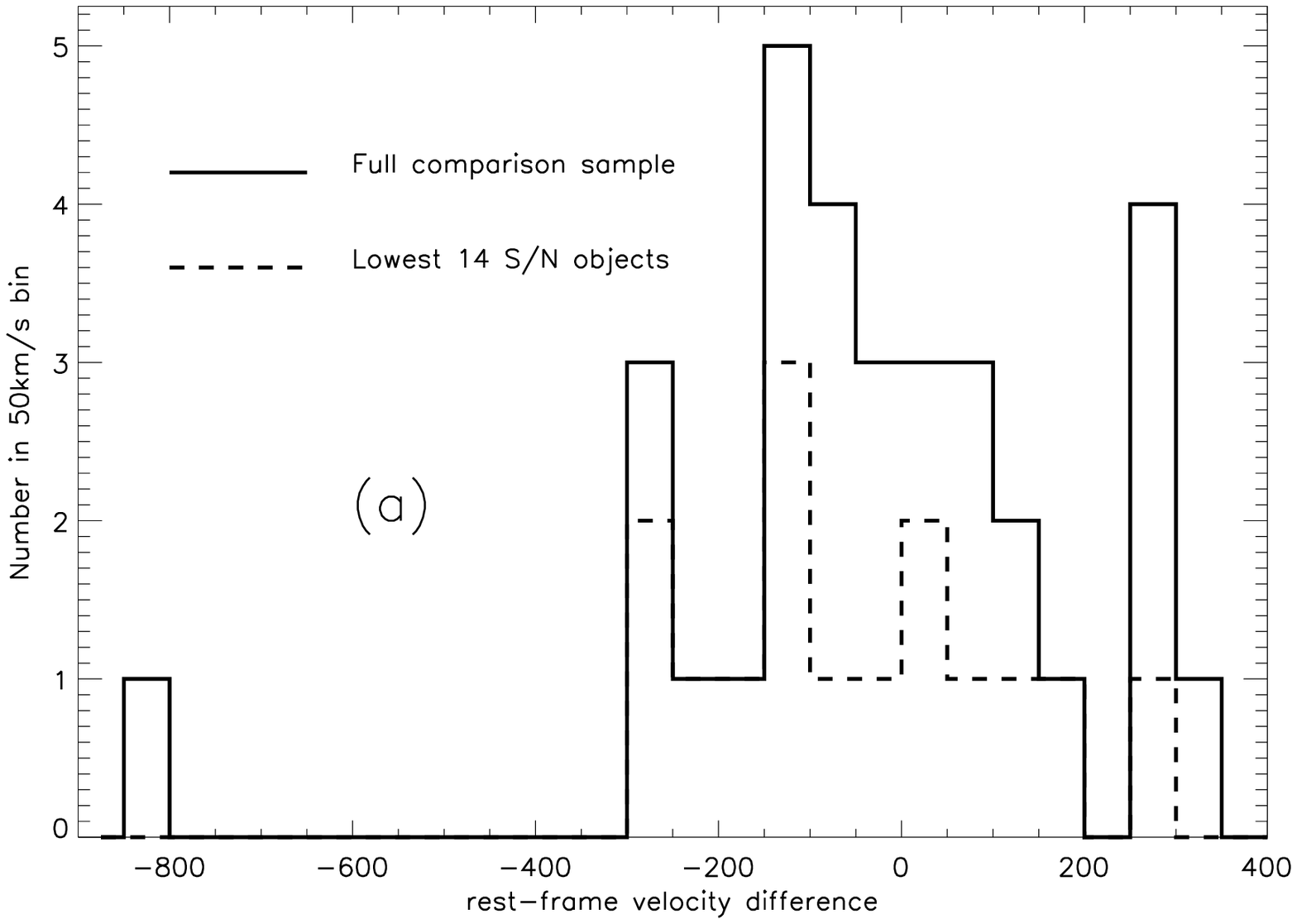,width=0.5\textwidth}\psfig{figure=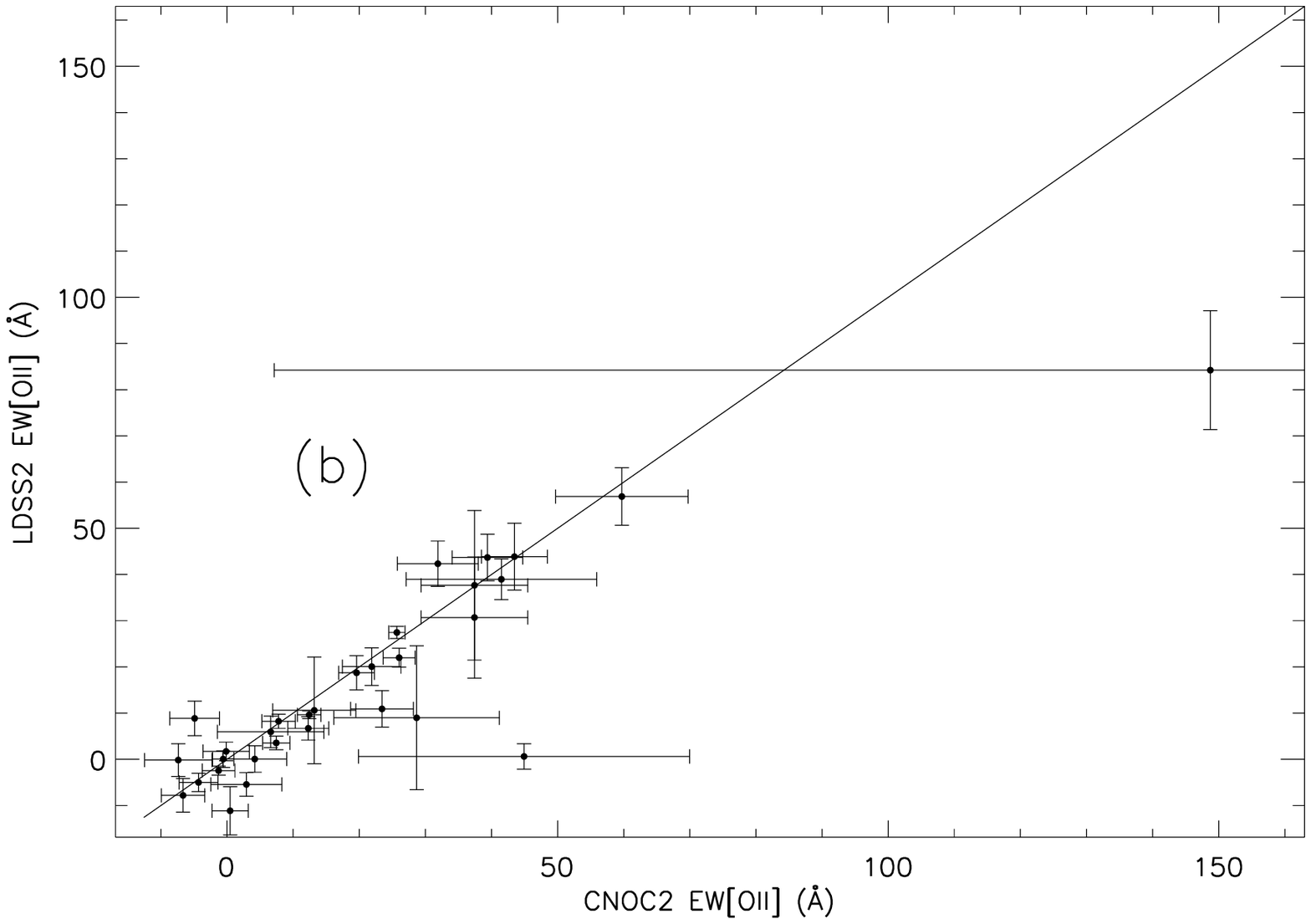,width=0.5\textwidth}}
\caption{A comparison of velocity and EW[OII] measurements in CNOC2 and
  LDSS2 measurements using the 31 galaxies with matched redshifts in
  the comparison sample. {\bf (a):} The distribution of rest-frame velocity 
offset, $(cz_{CNOC2} - cz_{LDSS2})/(1+z)$, in all galaxies (solid line) and the 14 galaxies with the lowest combined signal 
in the wavelength ranges $5300$\AA$-5530$\AA\ and $5645$\AA$-5820$\AA (dashed line, see text for more detail).
   and {\bf (b):} Comparison between
  rest-frame EW[\oii] measurements, for the galaxies in which redshifts
  and EW[\oii] have been obtained from LDSS2 spectra \emph{and} in the
  original CNOC2 catalogue.} 
\label{fig:Zoiicomp}
\end{figure*}

In Figure~\ref{fig:Zoiicomp} we compare the redshift
and EW[\oii] measurements obtained from the LDSS2
spectra and the CNOC2 spectra, for objects in common to both
surveys. It can be seen that the redshifts from the two surveys are
generally in good agreement. Out of 35 galaxies with redshifts from
both sources, only four of the redshifts are discrepant. Three of these are
faint objects ($B > 23.5$, $R_c > 21.5$).
We also examine the combined spectral signal (total counts, \emph{cts})
from the wavelength ranges $5300$\AA$-5530$\AA\ and
$5645$\AA$-5820$\AA\ which span the most efficient region of the
grism, eliminating the strong night sky lines (as used to examine targets for which no redshifts were measured in Appendix~\ref{sec:zincompl}). These three objects all fall in the low signal range ($cts < 1.5\times10^{4}$) along with 8 of the remaining 32 galaxies in the combined sample. We exclude the three discrepant redshifts for faint, low signal objects from the figure.
The other discrepant redshift is offset by $\sim
800$~km~s$^{-1}$ (rest-frame). Inspection shows that the LDSS2 redshift
provides a better fit to the emission lines. Of the remaining 31
galaxies in this sample, the standard deviation of the rest-frame
velocity offsets ($(cz_{CNOC2} - cz_{LDSS2})/(1+z)$), $\dv_{tot} =
175$~km~s$^{-1}$ with a mean value of only $-6$~km~s$^{-1}$. 
To check the dependence of rest-frame velocity offsets on the spectral 
counts, we also show the distribution for the 14 galaxies with $cts < 2.25\times10^{4}$ (dashed histogram). We find no correlation, amongst these remaining objects, between spectral signal strength and velocity offset.

Using duplicate observations \cite{Yee00} show that the random
rest-frame velocity errors in the CNOC2 redshifts is
$103$~km~s$^{-1}$. As the typical velocity difference in our
common-object sample is $175$~km~s$^{-1}$, we compute an approximate
rest-frame error on LDSS2 spectra of $\dv_{LDSS2} = \sqrt{175^2 -
  103^2} = 142$~km~s$^{-1}$. We expect that the velocity errors are
dominated by the CNOC2 astrometrical errors of $\lsim 1$\arcsec\
\citep{Yee00}. Indeed the errors computed using the common-object
sample are consistent with an rms astrometrical error of $\sim
0.25$\arcsec.  

In the right hand panel of Figure~\ref{fig:Zoiicomp}, we show that
measurements of EW[\oii] in LDSS2 spectra (measured using our code) are
also in excellent agreement with the original CNOC2 survey measurements
\citep{Whitaker04} within errors, showing that our measurements are
consistent with CNOC2. The code used for both sets of measurements is
essentially the same and tests show that they provide identical
results. 

\subsection{Group Membership}\label{sec:grpmem}

To obtain a consistently selected sample of galaxy groups, we
restricted ourselves to examining those groups which were pre-selected
from the sample of \citet{Carlberg01} and targetted with LDSS2 on
Magellan. This provides a sample of 26 targetted and serendipitous
\citet{Carlberg01} groups in the range $0.3 \leq z \leq 0.55$. We
derive from the virial theorem $r_{200} \sim 
\sigma(v){\left[11.5H_{75}(z)\right]}^{-1}$ where $\sigma(v)$ is the
velocity dispersion and $H_{75}(z)$ is the Hubble constant at the group
redshift z and $H_{75}(0) = 75$~km~s$^{-1}$~Mpc$^{-1}$. This means
that a group with $\sigma(v) = 300$~km~s$^{-1}$ has $r_{200} =
0.28h_{75}^{-1}$Mpc at $z=0.4$, with $r_{200}$ scaling with
$\sigma(v)$. The group finding algorithm of \citet{Carlberg01} was
tuned to identify dense, virialized groups of 3 members or more with
the goal of tracing the properties of the underlying dark matter
halos. With this in mind, a conservative linking-length
($0.33h_{75}^{-1}$Mpc) and trimming radius of 1.5$r_{200}$ in the
spatial axes of the groups were selected.  

Our objectives differ from those of \citet{Carlberg01} in that we wish
to understand the global properties of galaxies in groups, not only
those galaxies in the virialised core regions. In order to be more
representative of the loose group population, whilst retaining the
strict selection criteria of \citet{Carlberg01}, we choose to relax the
projected trimming radius for group members in our redefined group
sample. In effect, we find that relaxing this parameter allows the more
compact groups to retain the same membership whilst other groups gain
extra members.  

Although defining the contents of a group is a subjective problem,
there exist some tools which make the task easier. Perculiar motion of
galaxies moving in a gravitational potential artificially lengthens the
group along the line-of-sight direction in redshift space (the finger-of-God effect). From the virial theorem, one can make the assumption
that the projected spatial and line-of-sight dimensions of a group in
redshift space are approximately in constant proportion (in rest-frame 
coordinates). \citet{Eke04}
show that an axial ratio of $\sim 11$ for the length along the line of
sight relative to the projected spatial length
is most appropriate for a linking volume 
in a friends-of-friends algorithm. Interlopers in redshift space are
difficult to identify and eliminate, so we choose a conservative
line-of-sight trimming radius of twice the velocity
dispersion. 

We choose a small aspect ratio of $b=3.5$, to compute our trimming radius.
Although this will exceed $r_{200}$, it results in a stable
membership solution for all 26 groups and allows us to examine how
radial trends within the groups extend to the group outskirts and infall
regions (see Section~\ref{sec:fpstacked}). However most of our analysis will 
be conducted within a fixed metric group aperture. We note that these results 
are insensitive to the choice of a particular value of b, as large as 11.

\newpage

The algorithm for defining the final membership of each group in our
sample works as follows: 
\begin{itemize}
\item{The group is initially assumed to be located at the latest
    position in redshift space determined by
    \citet{Carlberg01}\footnote{Note: these positions have been updated
      since publication} with an initial observed-frame velocity
    dispersion, $\svo$, of 500~km~s$^{-1}$.} 

\item{The redshift range required for group membership is computed from
    equation~\ref{eqn:ZRange}, which limits membership in the
    line-of-sight direction to within twice the velocity
    dispersion:\newline
\begin{equation}
  \dz_{max} = 2\frac{\svo}{c}
\label{eqn:ZRange}
\end{equation}
}

\item{This is converted into a spatial distance, $\dth_{max}$, which
    corresponds to a redshift space distance related to the distance
    computed in the line-of-sight direction by the aspect ratio,
    $b$. $\dth_{max}$ is computed as:

\begin{align}
  \dth_{max} = 206265\arcsec.\left[\frac{\dr_{max}}{h_{75}^{-1}\mbox{Mpc}}\right].\left(\frac{D_{\theta}}{h_{75}^{-1}\mbox{Mpc}}\right)^{-1}
\label{eqn:phystoang}
\end{align}
where $D_{\theta}$, the angular diameter distance in
physical co-ordinates, is a function of
$z$, and
}

\begin{equation}
\dr_{max}  =  \frac{c \dz_{max}}{b (1+z) H_{75}(z)}   
\label{eqn:ProjAngRange}
\end{equation}
with $b = 3.5$.

\item{Group members are selected by applying the redshift and
    positional limits $\dz = |z-z_{group}| \leq \dz_{max}$ and the
    angular distance from the group centre, $\dth \leq \dth_{max}$.} 

\item{We recompute the observed velocity dispersion of the group,
    $\svo$, using the Gapper algorithm (equation~\ref{eqn:GapperVD})
    which is insensitive to outliers and thus gives an accurate
    estimate of the velocity dispersion for small groups
    \citep{Beers90}: \footnote{The multiplicative factor 1.135 corrects
      for the $2\sigma$ clipping of a gaussian velocity distribution.} 

\begin{equation}
  \svo = 1.135c \times \frac{\sqrt{\pi}}{n(n-1)} \sum_{i=1}^{n-1} w_{i}g_{i}
  \label{eqn:GapperVD}
\end{equation}
with $w_{i} = i(n-i)$ and $g_{i} = z_{i+1}-z_{i}$.
} 

\item{We also redefine the centre of the group, taking the luminosity
    weighted centroid in projected spatial coordinates and the mean
    redshift to be the new group centre.} 

\item{We then recompute the limiting redshift offset $\dz_{max}$ and
    positional offset $\dth_{max}$ using
    equations~\ref{eqn:ZRange}, \ref{eqn:phystoang}
    and~\ref{eqn:ProjAngRange}. The galaxies are then reassigned to the
    group as before.} 

\item{The whole process is repeated until a stable membership solution
    is reached. In all 26 groups such a stable solution is found, mostly
    within 2 iterations although the massive group 138 requires 4
    iterations. We finally compute the rest-frame velocity dispersion
    (equation~\ref{eqn:convrestVD}) and the intrinsic velocity
    dispersion. The latter is computed by combining the measurement
    errors of the component galaxies (equation
    ~\ref{eqn:combmeasvderr}) and removing in quadrature from the
    measured velocity dispersion (equation
    ~\ref{eqn:remmeasvderr}).\newline 

\begin{equation}
  \svr = \frac{\svo}{1+z}
  \label{eqn:convrestVD}
\end{equation}

\begin{equation}
  <\dv>^2 = \frac{1}{N}\sum_{i=1}^N\dv_i^2
\label{eqn:combmeasvderr}
\end{equation}
where $\dv = 142$~km~s$^{-1}$ (LDSS2) \newline
and $\dv = 103$~km~s$^{-1}$ (CNOC2) \newline

\begin{equation}
  \svi^2 = \svr^2 - <\dv>^2
\label{eqn:remmeasvderr}
\end{equation}
}

\end{itemize}

The combined velocity errors ($<\dv>$) in group 24 exceed the measured
velocity dispersion $\svr$ and so we compute an upper limit to the
intrinsic velocity dispersion using Monte-Carlo simulations. With
$\svi=119.6$~km~s$^{-1}$ in $15.87$ per cent of iterations a value of
$\svr$ less than the measured value is obtained from the simulations,
corresponding to the $1\sigma$ value of a one tailed gaussian. Errors on
$\svi$ are computed in all other groups using the Jackknife technique
\citep{Jackknife}. We note that in groups with few known members
(e.g. group 40) the true error on $\svi$ may be underestimated using this
technique.

\begin{figure}
\centerline{\psfig{figure=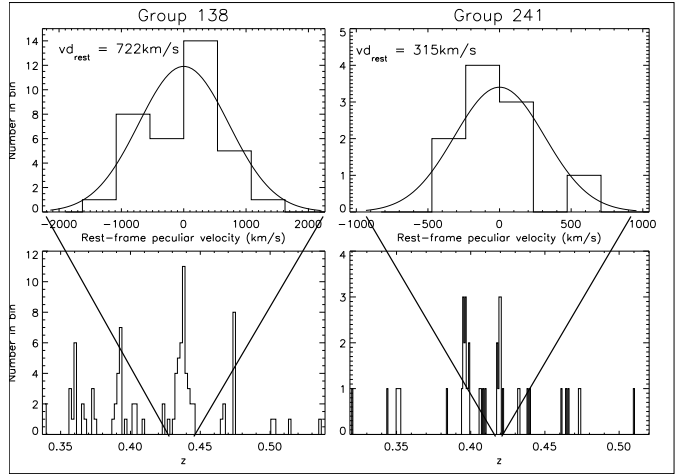,width=0.5\textwidth}}
\caption{Bottom: Redshift distributions in the regions of two groups;
  Top: Velocity distribution of the group members; The smooth line
  represents a gaussian with $\sigma = \svo$).} 
\label{fig:veldistrib} 
\end{figure}

In Figure~\ref{fig:veldistrib}, we show velocity histograms for two of
our groups and the redshift space clustering in the region of those
groups. The left-hand panels show the rich group 138, possibly even a
poor cluster, whilst the right-hand panels show the medium-sized group
241. The overplotted Gaussian of width $\svo$ provides a good envelope
to the distribution in both cases, which is typical of all 26
groups. We further note that there is no significant trend of velocity
dispersion with redshift in our sample indicating that we do not
preferentially select higher mass groups at higher redshift (where the
survey does not probe so far down the luminosity function). 

\begin{figure}
\centerline{\psfig{figure=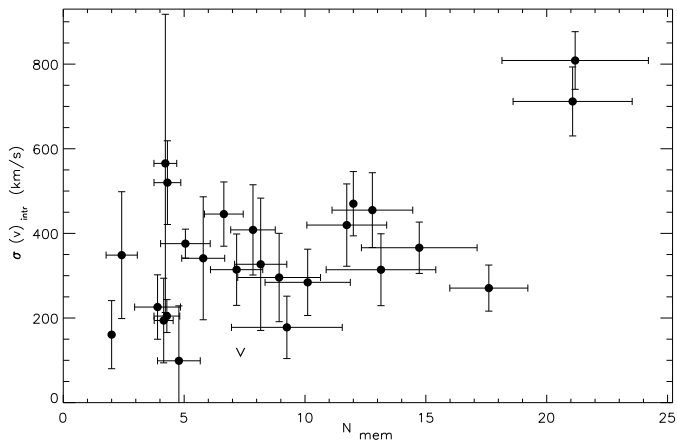,width=0.5\textwidth}}
\caption{Intrinsic velocity dispersion of galaxy groups, $\svi$ plotted
  against the number of galaxy members within 1~Mpc and brighter than
  \MBj\ $=-20$, corrected for incompleteness ($N_{mem}$). Errors on
  $N_{mem}$ represent the Poisson error on the number of candidate
  members estimated to lie in the group (see Section~\ref{sec:grpmem}
  for definition).} 
\label{fig:NMVD}
\end{figure}

Whilst group members lie some distance (up to $\sim4.5r_{200}$) from
the group centre and are included in the calculation of the velocity
dispersion, we limit our analysis to galaxies inside a
$1h_{75}^{-1}$Mpc projected radius at the group redshift ($\sim
3.5r_{200}$ for a group with $\svi = 300$~km~s$^{-1}$), where we
understand the completeness (see
Appendix~\ref{sec:completeness}). Rest-frame $B_J$ luminosities are
determined to minimise the K-corrections (the \Bj-band well matches the
$R_{c}$ selection band at the redshift of CNOC2 galaxies) and to allow
easy comparison with the properties of the 2dFGRS local galaxy
population (see Paper~II for more detail on determination of
K-corrections). Luminosities include a correction for Galactic
extinction on a patch-to-patch basis, computed by extrapolating from B
and V band extinction values obtained from NED \citep[][variation within each patch is negligible]{Schlegel98}. A luminosity \MBj\ $= -20$
approximates the magnitude limit, $R_{c}=22$ for most galaxies at
$z=0.55$. We compute $N_{mem}$, an estimate of the number of group
members brighter than \MBj\ $= -20$ and within
$1h_{75}^{-1}$Mpc projected distance from the group centre. $N_{mem}$
is the sum of the number of known members and candidate members
(the number of galaxies without redshifts which are predicted to lie at
the group redshift, and which would meet our luminosity and
radial cuts). Figure~\ref{fig:NMVD} shows $N_{mem}$ plotted against the
intrinsic velocity dispersion $\svi$ of each group. Any weak
correlation between $N_{mem}$ and $\svi$ is largely masked by a good
deal of scatter (excluding the two largest systems). This scatter might
be attributed to the variation in group structure and the difficulty of
obtaining accurate estimates of velocity dispersion with few
members. Limiting the membership to galaxies within $0.5h_{75}^{-1}$Mpc
projected radius does not reduce the scatter in this relationship. 

\section{Individual Group Analysis}\label{sec:Indivgrps}

\subsection{Basic Parameters}

\begin{table*}
\begin{minipage}{17cm}
\caption{Individual group properties: Luminosity-weighted
  group centroid positions in spatial coordinates (columns 2,3); mean redshift
  (column 4);
  velocity dispersion, $\svi$ (column 5); total number of galaxies in each group
  with a redshift, $N_{tot}$ (column 6); the number of members brighter than \MBj$ =
  -20$ and within $1h_{75}^{-1}$Mpc, weighted to account for
  incompleteness, $N_{mem}$ (column 7); the group morphology classification
  (column 8); the luminosity of the brightest group galaxy (within
  $1h_{75}^{-1}$Mpc of the centre), \MBj\ ($Br$, column 9); and the fraction of passive
  galaxies in the group, $\fp$ (brighter than \MBj$=-20.0$ and within
  $0.5h_{75}^{-1}Mpc$ of the iteratively determined group centre,
  column 10) The 
  computation of $\fp$ includes resampling to account for galaxies
  without redshifts in the region of each group (See
  Section~\ref{sec:sfgrpsvd}).} 
\label{tab:grps}
\begin{center}
{\scriptsize
\begin{tabular}{cccccccccc}
\hline\hline
Group & RA (J2000) & Dec (J2000) & z & $\svi$(~km~s$^{-1}$) & $N_{tot}$ & $N_{mem}$  & Class & $M_{B_j}(Br)$ & $\fp$\\
\hline
23 & 14:49:25.0 & +09:30:26 & 0.351 & 445 $\pm$ 75 & 13 & 6.64 & C & -21.84 & 0.67 \\
24 & 14:49:03.9 & +09:06:57 & 0.359 & $<$119.6 & 11 & 7.33 & C & -21.7 & 0.71 \\
25 & 14:49:40.8 & +09:13:43 & 0.361 & 470 $\pm$ 75 & 19 & 12.00 & C & -21.17 & 0.63 \\
27 & 14:49:49.9 & +09:06:38 & 0.372 & 348 $\pm$ 149 & 8 & 2.41 & L & -22.35 & 1.00 \\
28 & 14:50:22.8 & +09:01:13 & 0.372 & 160 $\pm$ 80 & 6 & 2.00 & C & -20.47 & 1.00 \\
31 & 14:49:12.9 & +09:10:09 & 0.392 & 565 $\pm$ 352 & 5 & 4.22 & L & -21.06 & 0.33 \\
32 & 14:50:01.3 & +08:55:57 & 0.394 & 519 $\pm$ 98 & 9 & 4.30 & L & -21.05 & 0.25 \\
33 & 14:49:36.8 & +09:28:48 & 0.406 & 194 $\pm$ 100 & 7 & 4.15 & L & -20.98 & 0.80 \\
34 & 14:48:43.8 & +08:52:02 & 0.465 & 408 $\pm$ 106 & 11 & 7.84 & L & -21.71 & 0.25 \\
37 & 14:49:29.1 & +09:05:33 & 0.471 & 419 $\pm$ 97 & 19 & 11.72 & C & -21.66 & 1.00 \\
38 & 14:49:27.8 & +08:58:12 & 0.510 & 808 $\pm$ 68 & 19 & 21.17 & M & -22.58 & 0.29 \\
39 & 14:49:23.2 & +09:30:24 & 0.536 & 454 $\pm$ 88 & 15 & 12.78 & C & -22.41 & 0.88 \\
40 & 14:49:20.2 & +08:54:57 & 0.542 & 177 $\pm$ 73 & 4 & 9.25 & U & -21.04 & 0.33 \\
129 & 21:51:03.3 & -05:42:23 & 0.317 & 225 $\pm$ 76 & 7 & 3.90 & C & -20.98 & 0.50 \\
132 & 21:50:25.8 & -05:40:53 & 0.359 & 375 $\pm$ 34 & 9 & 5.05 & L & -21.02 & 1.00 \\
133 & 21:50:48.2 & -05:37:33 & 0.373 & 204 $\pm$ 38 & 4 & 4.28 & U & -20.41 & 0.00 \\
134 & 21:50:24.8 & -05:41:29 & 0.392 & 284 $\pm$ 78 & 12 & 10.11 & L & -20.99 & 0.33 \\
137 & 21:50:37.7 & -05:29:18 & 0.425 & 314 $\pm$ 84 & 8 & 7.17 & C & -22.27 & 1.00 \\
138 & 21:50:48.2 & -05:39:53 & 0.437 & 711 $\pm$ 81 & 35 & 21.07 & M & -21.83 & 1.00 \\
139 & 21:50:25.8 & -05:50:20 & 0.439 & 314 $\pm$ 84 & 12 & 13.14 & L & -21.69 & 0.71 \\
140 & 21:50:41.4 & -05:28:38 & 0.465 & 98 $\pm$ 129 & 5 & 4.78 & C & -21.78 & 0.33 \\
227 & 02:26:29.5 & +00:12:08 & 0.363 & 341 $\pm$ 145 & 7 & 5.78 & L & -21.49 & 0.33 \\
228 & 02:25:05.7 & -00:07:12 & 0.384 & 326 $\pm$ 156 & 8 & 8.16 & C & -21.66 & 0.67 \\
232 & 02:25:50.7 & +00:52:17 & 0.396 & 366 $\pm$ 60 & 14 & 14.72 & C & -21.81 & 0.29 \\
241 & 02:25:58.7 & +00:54:21 & 0.419 & 295 $\pm$ 104 & 10 & 8.93 & L & -21.38 & 0.43 \\
244 & 02:25:45.0 & +01:07:30 & 0.470 & 270 $\pm$ 54 & 18 & 17.60 & C & -22.20 & 0.10 \\
\hline
\end{tabular}
}
\end{center}
\end{minipage}
\end{table*}

In Table~\ref{tab:grps} we present some of the fundamental properties
of our CNOC2 group sample including the velocity dispersion, $\svi$ and
the number of members of each group (both the total with measured
redshifts, $N_{tot}$, and the number brighter than \MBj$ = -20$ and
within $1h_{75}^{-1}$Mpc of the group centre, weighted to account for
incompleteness, $N_{mem}$). The group \emph{class} and parameter $\fp$
is defined in Section~\ref{sec:sfindgrps}. Groups closer to the low
redshift limit of $z=0.3$ are on average complete down to fainter
luminosities. Incompleteness over the full luminosity range is also a
function of the fraction of targetted objects in the group vicinity. 

\subsection{Star Formation in the CNOC2 groups}\label{sec:sfindgrps}

\begin{figure*}
\centerline{\psfig{figure=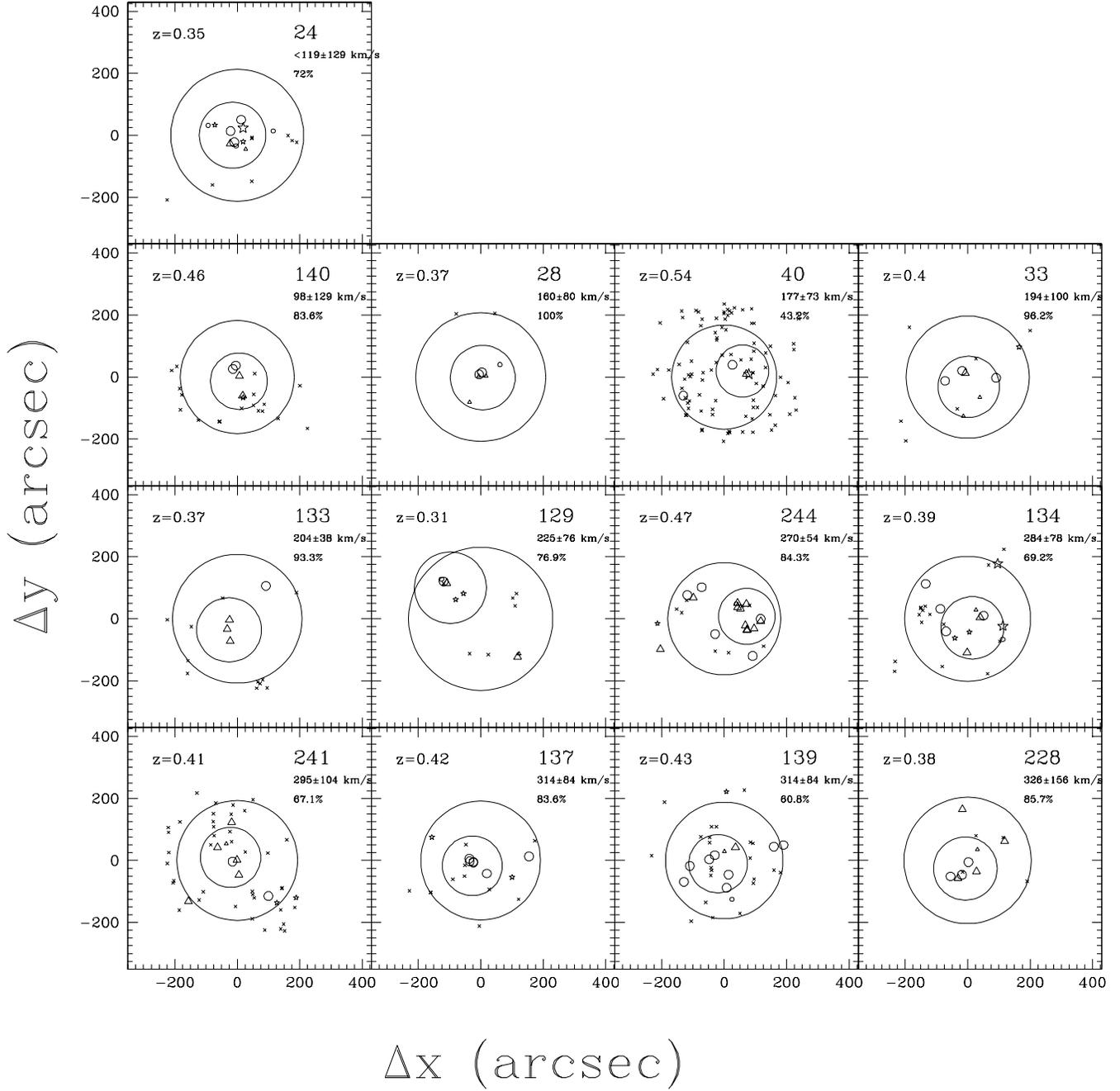,width=1.0\textwidth}}
\caption{The 13 groups with lowest $\svi$ Passive galaxies, star
    forming galaxies and highly star forming galaxies are
  represented by {\it circles, triangles} and {\it stars}
  respectively. Luminous members are represented by larger
  symbols than faint members. {\it Crosses} represent candidate
    luminous members. These are bright galaxies without measured
  redshifts and, while in some groups these will be mostly 
  foreground galaxies, in other groups (preferentially the
  more massive ones) a significant number may also be group
  members. The $1h_{75}^{-1}$Mpc radius of each group is also shown
  centred on the luminosity-weighted centre computed using all known
  members. We also show a $0.5h_{75}^{-1}$Mpc radius circle centred on
  an iteratively defined centre (see
  Section~\ref{sec:sfindgrps}). Definitions of these galaxy types can
  be found in the text (Section~\ref{sec:sfindgrps}). With each group,
  we also display the group redshift, velocity dispersion $\svi$ and
  completeness within $1h_{75}^{-1}$Mpc and brighter than the
  $R_{c}$-band magnitude required for the galaxy to have a luminosity
  \MBj$\leq -20$ should the galaxy be at the group
  redshift. Completeness is a function of redshift and the fraction of
  targetted objects in the group vicinity. Note that groups are ordered
  by velocity dispersion, $\svi$.} 
\label{fig:groupsA}
\end{figure*}

\begin{figure*}
\centerline{\psfig{figure=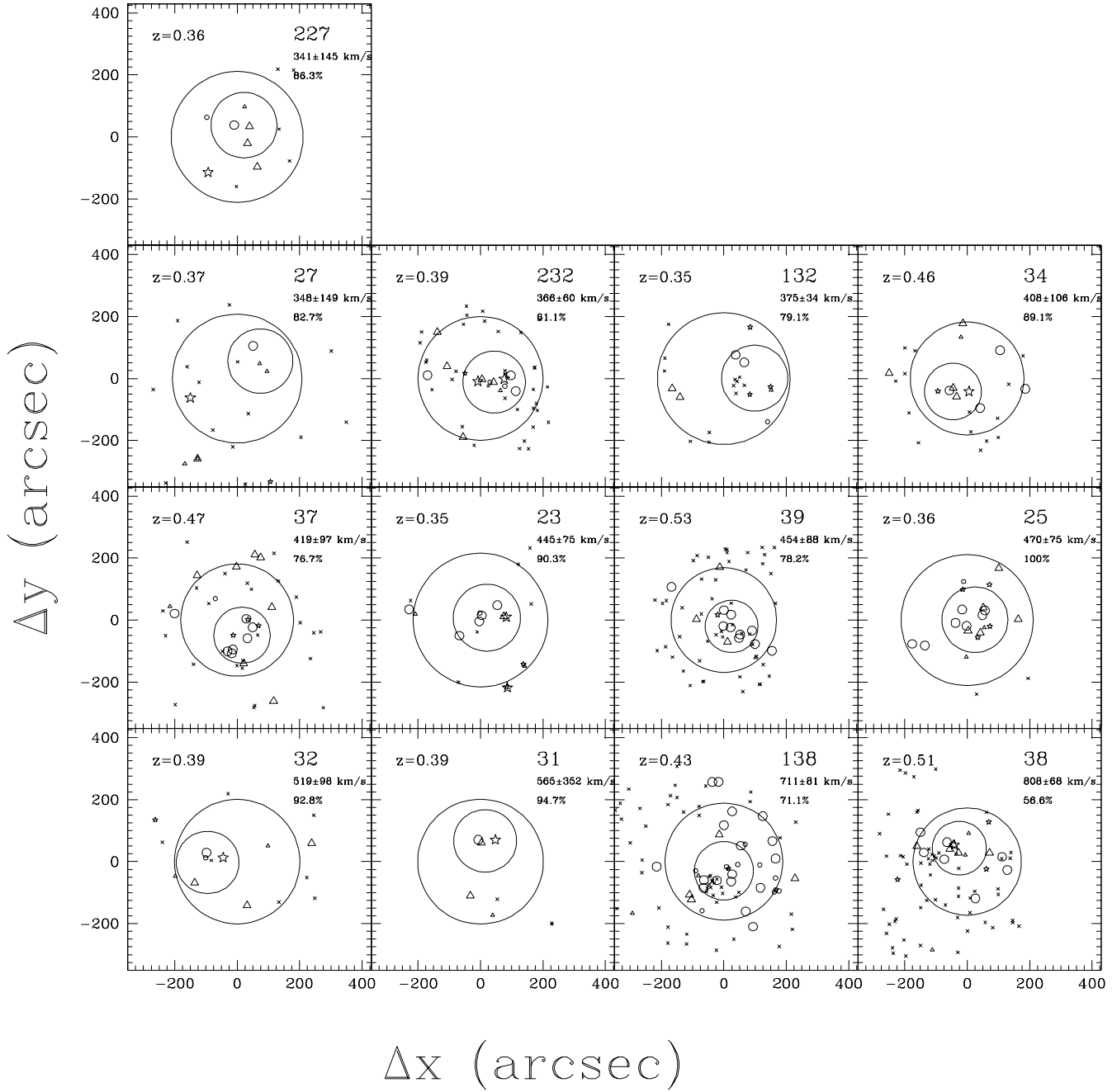,width=1.0\textwidth}}
\caption{As Figure~\ref{fig:groupsA} for the 13 groups with highest $\svi$.}
\label{fig:groupsB}
\end{figure*}

The local galaxy population shows a distinct bimodality in galaxy
properties. This is seen in galaxy colours
\citep{Strateva01,Blanton03,Baldry03} and in EW[\Halpha]
\citep{Balogh03}. In Section~\ref{sec:sfstacked} and Paper~II we shall
show that an artificial division in EW[OII] at 5\AA\ is sufficient to
reveal trends in the fraction of star-forming galaxies
(EW[OII]$\geq 5$\AA) and passive galaxies (EW[OII]$<5$\AA) in
the stacked group. We also create a third category of objects, the
highly star-forming galaxies, with EW[OII]$\geq 30$\AA. The spatial
distribution of these three types of objects in our 26 
groups reveals the connection between
star formation and the local environment of galaxies within each
group. Figures~\ref{fig:groupsA} and~\ref{fig:groupsB} show the spatial
distribution of these three types of galaxy in our 26 groups, ordered by their velocity
dispersion $\svi$. 
At our high redshift limit of $z=0.55$, galaxies brighter than
our magnitude limit $R_{c}=22$ possess luminosities of \MBj$\lsim
-20$. Therefore group members brighter than this luminosity
(luminous members) are represented by larger symbols, while
faint members (\MBj$> -20$) are represented by smaller
symbols. A limiting radius for our LDSS2 targetting is typically $\sim
240$\arcsec, which corresponds to $1h_{75}^{-1}$Mpc at the low redshift
end of our sample, $z=0.3$. We represent the $1h_{75}^{-1}$Mpc radius
centred on the luminosity-weighted centroid of all known members in
each group, with an overplotted circle. We also compute an iteratively
defined centre by throwing out galaxies beyond $1h_{75}^{-1}$Mpc and
recomputing the luminosity-weighted centroid. This is repeated twice,
reducing the radial limit to $0.75h_{75}^{-1}$Mpc and finally to
$0.5h_{75}^{-1}$Mpc. A $0.5h_{75}^{-1}$Mpc radius circle ($\sim
r_{200}$ for a $\svi = 500$~km~s$^{-1}$ group) centred on this
iteratively defined centre is also shown. Finally, it is pertinent to
recognise that a small fraction of galaxies in close proximity to the
group centre but without redshifts may also be members. Such galaxies
which would have luminosities \MBj$\leq -20$ should they lie at the
group redshift are overplotted as crosses. We refer to  these galaxies
as candidate luminous members.

As can be seen in Figures~\ref{fig:groupsA} and~\ref{fig:groupsB},
groups come in all shapes and sizes. Judgement about global group
properties is reserved for the stacked group which smoothes out the
peculiarities of each individual group. This is covered in
Section~\ref{sec:stackedgrp} and Paper~II. Some general trends are
nonetheless readily apparent in Figures~\ref{fig:groupsA}
and~\ref{fig:groupsB}: 
\begin{itemize}
\item{Some of the groups show clumps of either passive or star
    forming galaxies. The clumps of passive galaxies would be expected
    from the star formation - density relation
    \citep{Balogh03}. However, the clumps of star forming galaxies
    (seen for example in groups 244, 133, 232) could represent a part
    of the group system which had not been influenced by the group
    environment at the redshift of these groups.} 
\item{Passive galaxies are not restricted to the most centrally concentrated, isolated groups and the most massive groups. They are also present in loose and filamentary groups.}
\end{itemize}


\subsubsection{Trends with group concentration}\label{sec:sfgrpsconc}
\begin{figure}
\centerline{\psfig{figure=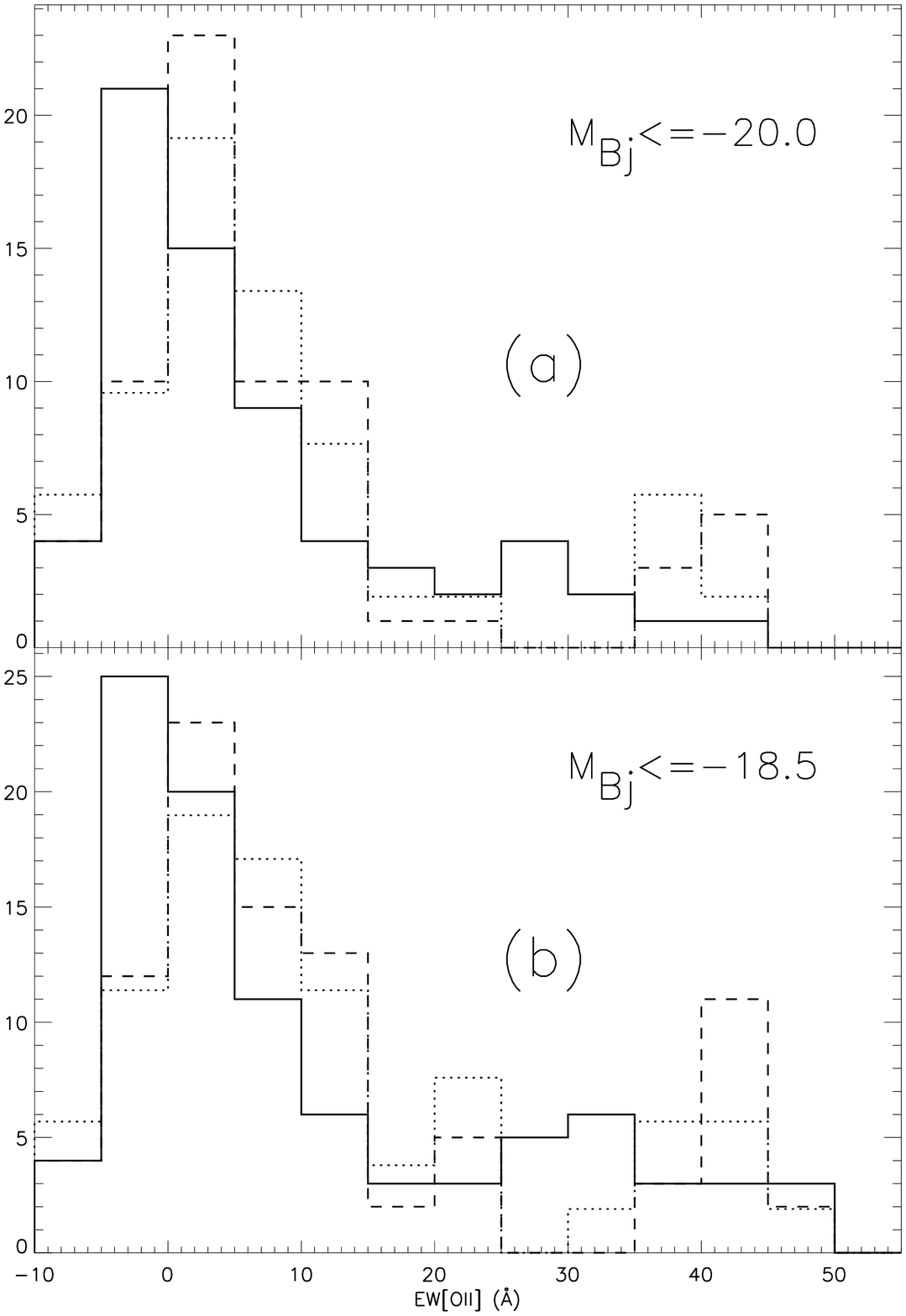,width=0.5\textwidth}}
\caption{{\bf (a)} The distribution of EW[OII] for the C-group galaxies (solid histogram), L-group galaxies (dotted histogram, renormalized to match the number of C-group galaxies) and the L-group galaxies, resampled to match the luminosity distribution of galaxies in the concentrated groups (dashed histogram). The sample is limited to galaxies brighter than \MBj$\leq -20.0$ and  to within $0.5h_{75}^{-1}Mpc$ of the iteratively determined group centre (smaller circles in Figures~\ref{fig:groupsA} and~\ref{fig:groupsB}). {\bf (b)} The same as above, except incorporating all galaxies down to the fainter limit of \MBj$\leq -18.5$.}
\label{fig:concgrpsOII}
\end{figure}
Figures~\ref{fig:groupsA} and~\ref{fig:groupsB} show the wide variety
of group structures present in the intermediate-age Universe, from low
$\svi$ compact groups (e.g. groups 24 and 28), and more massive groups
spanning the gap to poor-clusters (groups 38 and 138) to elongated
structures (group 232) and loose groups (e.g. groups 33 and
34). Incompleteness makes the application of any strict quantifiable
compactness parameter to the groups difficult. However, for the purpose
of studying basic trends in galaxy properties with the overall group
structure, we divide the groups into three qualitative but readily
identifiable catagories, hereafter known as the group
class. These categories are concentrated (C),
loose (L) and massive (M). Groups which cannot be
classified are labelled unclassifiable, (U). The more massive
groups 38 and 138 are class M, identifying the more cluster-like
environment in these richer groups ($\svi \ge 600$~km~s$^{-1}$ and $N_{mem} \ge 20$). Groups in
which galaxies are located in dense clumps (and therefore biased
towards higher density in, for example, the morphology-density
relation) are classified as class C. More loosely clustered groups are
classified as class L and groups 40 and 133 are not classified at all
(class U) as each possesses only 4 confirmed members with no structure
readily apparent. The class of each group is shown in
Table~\ref{tab:grps}. 

We wish to see how the concentration of group galaxies (i.e. the group
structure or state of virialisation) might affect the underlying star
forming properties of the group galaxies.  
In Figure~\ref{fig:concgrpsOII}, we show the distributions of EW[OII]
in the C-group galaxies (solid histogram) and L-group galaxies (dotted
histogram; renormalised to match the number of C-group
galaxies). Galaxies are limited to within $0.5h_{75}^{-1}$Mpc of the
iteratively determined group centre (smaller circles in
Figures~\ref{fig:groupsA} and~\ref{fig:groupsB}). In the top panel,
galaxies are limited to \MBj$\leq -20.0$ and in the bottom panel we
include all group galaxies down to \MBj$\leq -18.5$. The EW[OII]
distribution of the L-group galaxy population is well matched to the
C-group galaxy population (as confirmed by a K-S test). We ensure that
this result is unchanged when the luminosity distributions of the two
samples are exactly matched, using a resampling technique. This
involves choosing an L-group galaxy closely matching the luminosity of
each C-group galaxy, which results in some L-group galaxies being
chosen more than once and others not at all. The dashed histogram in
Figure~\ref{fig:concgrpsOII} represents the resampled L-group galaxy
EW[OII] distribution. The resampling process has negligibly altered the
L-group galaxy EW[OII] distribution which is still therefore consistent
with the C-group galaxy population. 

Dividing the galaxies once more on
their EW[OII] into passive galaxies (EW[OII]$<5$\AA), star
  forming galaxies ($5$\AA$ \leq$EW[OII]$\leq 30$\AA) and highly
  star forming galaxies (EW[OII]$\geq 30$\AA), we can take the
fraction of galaxies in each category directly from
Figure~\ref{fig:concgrpsOII}. Table~\ref{tab:fracconcloose} shows the
fraction of passive galaxies, $\fp$, and the fraction of highly
star-forming galaxies, $\fhsf$, in both the C-group galaxy population,
and the L-group galaxy population (raw and resampled) with both
luminosity limits applied. We note that the small differences between
the two classes (less passive and more highly star forming galaxies in 
the loose groups) are not significant, and that these figures should not be 
overinterpreted as the group classification is uncertain. 
\begin{table}
\caption{The fractions of passive ($\fp$, EW[OII]$<5$\AA) and highly star forming ($\fhsf$, EW[OII]$\geq30$\AA) galaxies in C-group galaxies (concentrated into dense clumps), L-group galaxies (looser structure) and resampled L-group galaxies (to match the luminosity distribution of C-group galaxies). Galaxy populations are defined within 0.5$h_{75}^{-1}$Mpc of the iteratively determined group centre and down to \MBj$=-20.0$ (1) and \MBj$=-18.5$ (2). Statistical errors are computed using the Jackknife technique.}
\label{tab:fracconcloose}
\begin{center}
\begin{tabular}{ccccc}
\hline\hline
Class & $\fp$ (1) & $\fhsf$ (1) & $\fp$ (2) & $\fhsf$ (2)\\ 
\hline
C &  $61 \pm 6\%$ & $6 \pm 6\%$ & $53 \pm 5\%$ & $16 \pm 7\%$\\
L & $51 \pm 9\%$ & $11 \pm 5\%$ & $39 \pm 7\%$ & $18 \pm 4\%$\\
Resampled L & $55 \pm 6\%$ & $12 \pm 6\%$ & $42 \pm 5\%$ & $20 \pm 5\%$\\
\hline
\end{tabular}
\end{center}
\end{table}

\subsubsection{Trends with group velocity dispersion}\label{sec:sfgrpsvd}

\begin{figure}
\centerline{\psfig{figure=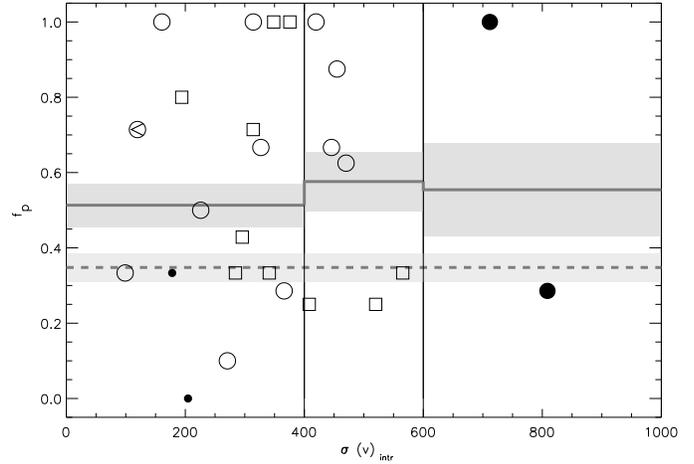,width=0.5\textwidth}}
\caption{The fraction of passive galaxies $\fp$ (brighter than
  \MBj$\leq -20.0$ and within  $0.5h_{75}^{-1}$Mpc of the iteratively
  determined group centre), in each group as a function of the group
  velocity dispersion, $\svi$. The galaxies in each group without
  redshift measurement (candidate members) are resampled from the
  measured-z galaxies in the same region of sky, on the basis of their
  luminosity and ($B-R_c$) colour. Groups are also keyed on the
  allocated group-class with {\it open circles} to represent concentrated
  groups, {\it squares} to represent loose groups, {\it large filled circles} to
  represent massive groups and {\it small filled circles} to represent
  unclassified groups. The arrow indicates an upper limit on the
  velocity dispersion of group 28. We also show the value of $\fp$ in
  combined groups for three bins of $\svi$ {\it (solid line)} and for the field
  {\it (dotted line)}. The {\it shaded area} represents the $1\sigma$ error on these values
  computed using the Jackknife technique. There is no clear trend of
  $\fp$ with $\svi$ in groups for galaxies brighter than \MBj$\leq
  -20.0$.} 
\label{fig:fp_VD}  
\end{figure}
The value of $\fp$ in each group is computed for all known group
galaxies within $0.5h_{75}^{-1}$Mpc of the iteratively determined group
centre and brighter than \MBj$=-20.0$. We also include candidate
members in the sample (see Section~\ref{sec:grpmem} for
definition). For each candidate member we assign the properties
(redshift, EW[OII]) of a galaxy with a measured redshift and a similar
($B-R_C$) colour and $R_C$ 
magnitude (referred to as a measured-z
  galaxy). This galaxy must also lie in the projected region of the
group, and a galaxy 0.5~mags different in $R_C$ is considered an
equally good match to one 0.25~mags different in ($B-R_C$)
colour. In this way a high redshift group (where a \MBj$=-20.0$ galaxy
has $R_C \sim 22.0$) is evenly sampled in luminosity despite
incompleteness at faint magnitudes. Nonetheless, we find that every
group $\fp$ is consistent with the value computed from known
members only. Values of $\fp$ for each group are also shown in
Table~\ref{tab:grps}. $\fp$ is plotted against group velocity
dispersion, $\svi$ in Figure~\ref{fig:fp_VD}. Groups are also keyed on
their class. There is no clear trend of $\fp$ with $\svi$ and the
scatter in the individual values is large because of the small number
of galaxies in each group. In order to improve the statistics in this
plot, we bin the groups in velocity dispersion. The average $\fp$ in
each bin shows no systematic evidence for variation with $\svi$. In
general, there is little trend of $\fp$ with group class. Only for
systems with $\svi > 400$~km~s$^{-1}$ velocity dispersion is there a
suggestion that the concentrated groups have higher $\fp$ than the
loose systems. We note that $\fp$ also shows no correlation with group
redshift within the sample.  

\section{The Stacked Group}\label{sec:stackedgrp}
We have shown that there are no significant differences between groups
of different class or velocity dispersion, so we now consider their
properties when combined. We co-add the data from our 26 groups to form
a stacked group. We then possess the statistical tools necessary
to investigate the global properties of group galaxies and to make a
comparison with the field at intermediate redshift ($0.3 \leq z \leq
0.55$). The stacked group sample contains a total of 282 galaxies above
our magnitude limit of $R_{C}=22.0$. Each galaxy is weighted by a
combined selection weight $W_C$ to correct for the stacked group
selection functions. These are well understood and are discussed in
detail in Appendix~\ref{sec:completeness}. However, we only trust the
weights when applied to a large sample as incompleteness varies from
group to group. Whilst the application of this weight is strictly
correct, we find the results from Section~\ref{sec:sfstacked} are
consistent with and without the galaxy weightings. 

We also define a field sample to contrast with our stacked group
sample. Field galaxies are defined to include all galaxies not
  associated with the \citet{Carlberg01} group sample, with $0.3 \leq
z \leq 0.55$ and within 240\arcsec\ of the targetted group centre
(where the radial selection function is well defined). The field sample
contains a total of 334 galaxies above our magnitude limit of
$R_{C}=22.0$.  

\subsection{The properties of stacked group galaxies}\label{sec:sfstacked}
In this section we investigate how galaxy properties in the stacked
group differ from the field population.  

\subsubsection{The fraction of passive galaxies in the stacked group}\label{sec:fpstacked}
\begin{figure}
\centerline{\psfig{figure=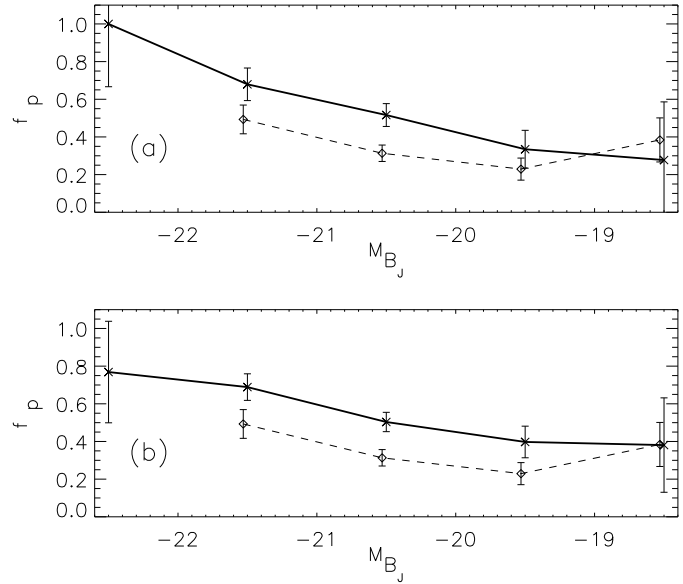,width=0.5\textwidth}}
\caption{{\bf (a):} The fraction of passive galaxies, $\fp$, in the
stacked group within $0.5h_{75}^{-1}$Mpc of the iteratively determined
group centre (solid line) and the field (dashed line) as a function of
galaxy luminosity, \MBj. The field symbols are offset slightly in
luminosity for clarity. Statistical errors on $\fp$ are computed using
a Jackknife method. {\bf (b):} The same as above, but including
all galaxies within $1h_{75}^{-1}$Mpc of the original centroid
computed in Section~\ref{sec:grpmem} (the luminosity-weighted centre
of all confirmed group members).}
\label{fig:fplum}
\end{figure}

In Figure~\ref{fig:fplum} we show how the fraction of passive galaxies,
$\fp$, depends upon galaxy \Bj-band luminosity in the stacked group and
the field. This should be distinguished from $\fp$ computed for each
individual group in Section~\ref{sec:Indivgrps} and
Figure~\ref{fig:fp_VD}. In the top panel of Figure~\ref{fig:fplum}, we
limit the group sample to galaxies within $0.5h_{75}^{-1}$Mpc of the
iteratively determined group centre (corresponding to $r_{200}$ for
$\svi \sim 500$~km~s$^{-1}$ groups). There is a clear enhancement of
$\fp$ in the group galaxies with respect to the field, especially in
the luminosity range $-22.0 \leq $\MBj$\leq -19.0$. In the lower panel,
we include all galaxies within $1h_{75}^{-1}$Mpc of the original
centroid computed in Section~\ref{sec:grpmem} (the luminosity-weighted
centre of all confirmed group members). There is no significant
change in $\fp$ as a function of luminosity when the radial constraint
is relaxed, apart from an improvement in the sample
statistics. Combining all galaxies within the luminosity range $-22.5
\leq$\MBj$\leq -18.5$, the enhancement in groups of $\fp$ is better
than $3\sigma$ significance, and this trend is still evident if the
massive groups 38 and 138 are excluded from the sample. However we note
that brighter than \MBj$=-21.0$ the value of $\fp$ is only enhanced in
groups of higher velocity dispersion ($\svi \gsim
400$~km~s$^{-1}$). During the remainder of our analysis we limit the
galaxy sample to those within $1h_{75}^{-1}$Mpc from the
luminosity-weighted centre of all confirmed group members. 

\begin{figure}
\centerline{\psfig{figure=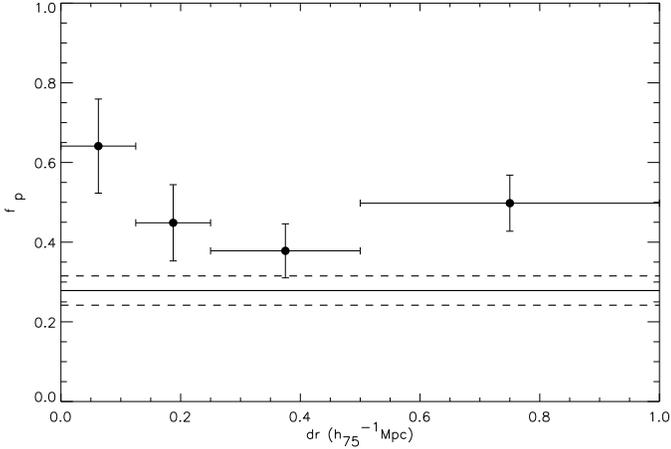,width=0.5\textwidth}}
\caption{$\fp$ (\MBj$\leq -18.5$) in the stacked group as a function of
  physical distance from the group centre where the group centre is
  defined to be the luminosity-weighted centre of all confirmed group
  members. 
The horizontal solid line and dashed lines represent the field $\fp$
and $\pm 1\sigma$ uncertainties respectively
(The luminosity limit is deeper than in Figure~\ref{fig:fp_VD} and so $\fp$ is reduced).
All errors on $\fp$ are computed using the Jackknife method.}  
\label{fig:fprad}
\end{figure}

\begin{figure}
\centerline{\psfig{figure=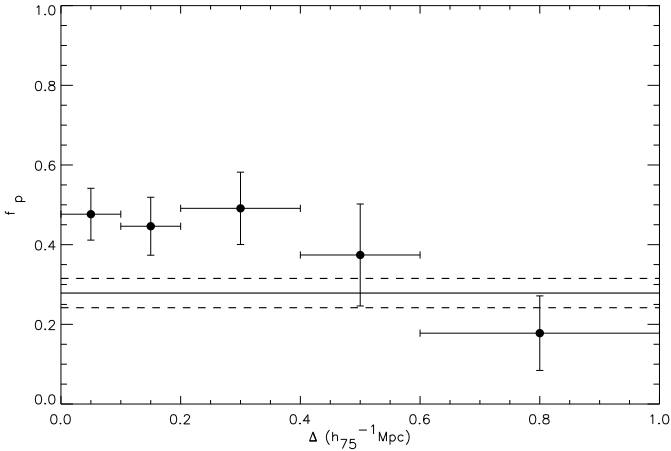,width=0.5\textwidth}}
\caption{$\fp$ (\MBj$\leq -18.5$) in the stacked group as a function of the projected distance to the nearest neighbour ($\Delta h_{75}^{-1}$~Mpc). 
The horizontal solid line and dashed lines represent the field $\fp$
and $\pm 1\sigma$ uncertainties respectively. All errors on $\fp$ are
computed using the Jackknife method. 
We find a correlation between $\fp$ and $\Delta$ where group galaxies
located in more overdense regions ($\Delta \lsim 0.4h_{75}^{-1}$~Mpc)
have an enhanced likelihood of being passive. 
} 
\label{fig:fpden}
\end{figure}

In the top panel of Figure~\ref{fig:fprad}, we plot $\fp$ as a
function of group-centric radius $\dr$ for all group galaxies
brighter than \MBj$= -18.5$. We have used a physical distance scale 
since virial radii in groups are uncertain
due to the low number of galaxies and often irregular group
morphologies. The horizontal solid line and dashed lines represent the
field $\fp$ and $\pm 1\sigma$ uncertainties respectively. There is a weak 
trend of $\fp$ with $\dr$ and a significant enhancement
over the field is seen out to $1.0h_{75}^{-1}$~Mpc. 
The lack of strong trend is unsurprising: the groups are
not relaxed, spherical systems and often have multiple concentrations within 
the fiducial $1.0h_{75}^{-1}$~Mpc radius (see Figures~\ref{fig:groupsA} and
~\ref{fig:groupsB}).  To see if any trend becomes more obvious when the
group is centred on the BGG (Brightest Group Galaxy, chosen from
within 1$h_{75}^{-1}$Mpc of the luminosity-weighted group centre), we
recalculate the radial trend placing the group centre on the BGG.
This enhances $\fp$ in the innermost bin, but does not otherwise strengthen
the radial gradient. In Figure~\ref{fig:fpden} we show the dependence of $\fp$ on 
the distance to the nearest confirmed group galaxy. This measure of density has the 
advantage that it can be applied to systems with few members and irregular spatial 
distribution. 
In this plot, $\fp$ is significantly reduced at large separations, 
indicating that the high passive fraction seen in Figure~\ref{fig:fpden} at large $\dr$ is related to 
the clumpy distribution of group galaxies. 
We note, however that there is no dependence on separation less than $\sim 0.4h_{75}^{-1}$~Mpc.

\subsubsection{Environmental dependencies in the luminosity function of galaxies}\label{sec:CNOCenvLumFn}

\begin{figure}
\centerline{\psfig{figure=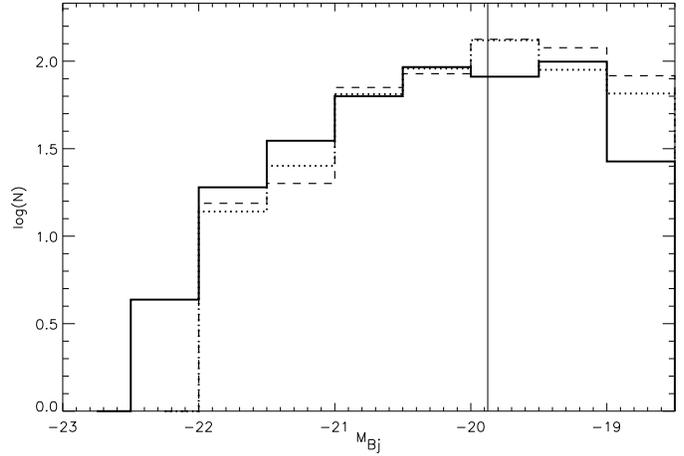,width=0.5\textwidth}}
\caption{The luminosity functions for group (solid line),
field (dotted line) and redshift-weighted field (dashed line) galaxies
weighted by selection, $W_{C}$ and within our redshift range ($0.3
\leq z \leq 0.55$). The vertical line represents the approximate
luminosity limits of the sample at the redshift limit $z=0.55$.}
\label{fig:lumfns_env}
\end{figure}

In Figure~\ref{fig:lumfns_env} we show how the
luminosity function of the stacked group (solid histogram) compares
with that of the field (dashed histogram) over the same redshift range
($0.3\leq z \leq 0.55$). For a full discussion of evolution
in the CNOC2 field luminosity function, see \citet{Lin99}; here we concentrate
on the comparison with the groups. The field
luminosity function has been scaled to match the number of galaxies seen in
the group luminosity function over the range $-21.0 \leq$\MBj$\leq
-20.0$. The vertical line represents the luminosity limit
of galaxies where $R_{c} = 22.0$ at $z=0.55$ (\MBj$=-19.75$), our high
redshift limit, for galaxies with mean
K-corrections. At the low redshift limit $z=0.3$ the corresponding luminosity limit is
\MBj$=-17.93$. For the galaxies with the largest K-corrections, these
limits become \MBj$=-18.06$ at $z=0.3$ and \MBj$=-20.07$ at
$z=0.55$. Figure~\ref{fig:lumfns_env} suggests that there may be a
small excess of bright (\MBj$\leq -21.0$) galaxies in the groups with 
respect to the field.  

A concern with this comparison is that the field
galaxies have a different redshift distribution from the group galaxies.
This is a particular concern at the faint end of the luminosity function
where the selection effect has the greatest impact.
In order to see if this can account for the difference between the field
and group luminosity functions, we apply an additional weighting to the
field galaxies to match the redshift distribution of the group
galaxies. The luminosity functions of the redshift-weighted
field is shown as a dotted line Figure~\ref{fig:lumfns_env}.
Within our luminosity range the field 
luminosity function remains relatively
unaltered by the redshift weighting. The comparison with the group luminosity function
is qualitatively unchanged. 

Figure~\ref{fig:lumfns_env} suggests that there is an excess of bright
galaxies and a deficit of faint galaxies in the groups compared to the field.
To investigate the statistical significance of this apparent excess of
bright galaxies in groups, we split the luminosity functions into three
bins of luminosity: bright (\MBj$\leq -21.0$), control
($-21.0 \leq$\MBj$\leq -20.0$; which encompasses the local value of
$\Mstar \sim -20.3$ from \nocite{Norberg02}{Norberg} {et~al.} 2002, 
corrected to $H_\circ=75$~km~s$^{-1}\Mpc^{-1}$) and faint ($-20.0
\leq$\MBj$\leq -18.5$). We prefer this approach over fitting a 
Schechter function to the luminosity function, since it avoids degeneracy
between the cut-off parameter, $\Mstar$, and the faint-end slope parameter,
$\alpha$. We define two new indices to measure the
relative abundances of galaxies in each of these luminosity bins. The
bright-to-faint galaxy ratio, $\BF$, is defined as the logarithm of
the ratio of galaxies in the bright bin to that in the faint bin. The
bright-to-control galaxy ratio, $\BC$ is similarly the logarithm of
the ratio of galaxies in the bright bin to that in the control
bin. The values of $\BC$ and $\BF$ in the field are shown in
Table~\ref{tab:BCBFField}.  $\BC$ only utilises data for which galaxies
are brighter than the magnitude limit at all redshifts, while $\BF$
spans a wider range of luminosity (and thus has more
leverage to measure changes in the luminosity function) but requires
us to correct by the redshift weighting (ie., to compare with the dotted 
line in Fig.~\ref{fig:lumfns_env}). We estimate the significance of 
the difference in the field and group ratios by bootstrap resampling 
two mock samples using the field data-set alone, and comparing
the difference between the two resampled values.  This indicates the 
likelihood that the observed difference between the group
and the field occurs by random sampling from the same underlying distibution.

A comparison between the group and field luminosity functions can now
be made by computing $\BC$ and $\BF$. The result of this process is 
summarised in Table~\ref{tab:BCBFField}. We compute values of $\BC =
-0.425$ for the group galaxy population and $\BC = -0.600$ for the
field galaxy populations, indicating a larger fraction of bright
galaxies in the groups. We compute a difference between the group and field of 
$\dBC = +0.175$; however, this result has low significance. The computed significance 
level for the difference $\dBC$ is $\SBC = 85.6$ per cent
(there is a $\sim$15 per cent chance that the luminosity functions are the same).
For $BF$ we find $\BF = -0.551$ in the group galaxy
population and $\BF = -0.976$ in the redshift-weighted field
population, again indicating a much larger ratio of bright to faint
galaxies in the groups than in the field ($\dBF = +0.425$). 
In this case, the difference has much higher statistical significance,
$\SBF = 99$ per cent
(we can be 99 per cent confident that the difference is real).\\

\begin{figure}
\centerline{\psfig{figure=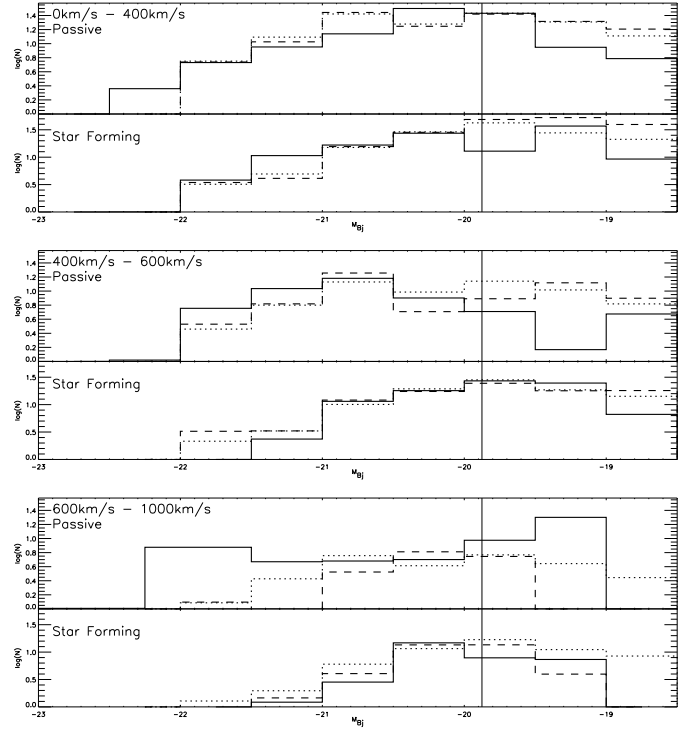,width=0.5\textwidth}}
\caption{As Figure~\ref{fig:lumfns_env} but with group galaxies split
  into three bins of velocity dispersion, $\svi$ going from low $\svi$
  ({\it top panels}) to high $\svi$ ({\it bottom panels}); and with
  star forming and passive galaxies 
shown separately in the upper and lower panels of each plot.} 
\label{fig:lumfns_bysfandvd}
\end{figure}

\begin{figure}
\centerline{\psfig{figure=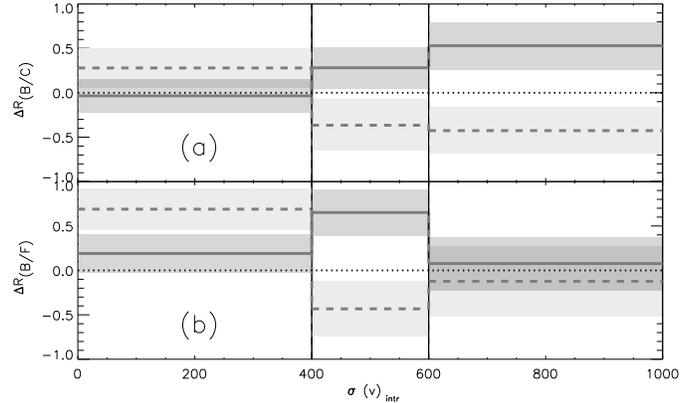,width=0.5\textwidth}}
\caption{The dependence on group velocity dispersion, $\svi$, of group luminosity function parameters, normalised by the field, as computed in Table~\ref{tab:BCBF}. {\bf (a)} The ratio of bright to control galaxies, $\dBC$ and {\bf (b)} The ratio of bright to faint galaxies, $\dBF$. The solid line represents the passive population, whilst the dashed line represents the star forming population of galaxies. The shaded areas represent the 1-$\sigma$ errors on the parameter, computed using a Monte Carlo technique. These are light grey for the star forming galaxies, darker grey for the passive galaxies, and darker still where the 2 regions overlap. A value of $\dBC$ or $\dBF = 0$ infers that there is no difference between group and field luminosity functions (dotted line).} 
\label{fig:dBCBFsig}
\end{figure}

Table~\ref{tab:BCBF} explores whether the field and group luminosity 
functions are significantly different for passive and actively star forming types, and whether
they differ as a function of group velocity dispersion.  Within the
uncertainty of the small sample size, the difference
in $\BF$ appears to be evident in both the star forming and passive populations.
While the excess of the brightest passive galaxies appears to be most prevalent
in the groups with highest velocity dispersion, the depression in the
faint galaxy population is most evident in the lower velocity dispersion
groups. Only in low velocity dispersion groups is a population of bright star forming galaxies 
still common. The changing shape of the luminosity function is shown in Figure~\ref{fig:lumfns_bysfandvd}. Figure~\ref{fig:dBCBFsig} shows us the dependence on $\svi$ of the parameters $\dBC$ and $\dBF$, graphically depicting the information in Table~\ref{tab:BCBF}. What is particularly evident in this figure is that whilst the bright, passive galaxies become more important in the high velocity dispersion groups (relative to $\Mstar$ galaxies), the bright star forming galaxies become less important. This suggests a bright star forming galaxy population in dense environments may contribute significantly to the formation of the most massive passive galaxies, already highly prevalent in these systems by $z \sim 0.5$.  

\begin{table}
\caption{{\bf Field Luminosity Function Properties:} The
  bright-to-control ($\BC$) and bright-to-faint ($\BF$) ratios of the
  galaxy populations in the CNOC2 field:} 
\label{tab:BCBFField}
\begin{center}
{\scriptsize
\begin{tabular}{cccccc}
\hline\hline
\multicolumn{3}{c}{$\BC$} & \multicolumn{3}{c}{$\BF$}  \\
All & Passive & Star & All & Passive & Star  \\
 &  & Forming & &  & Forming \\
\hline
-0.600 & -0.401 & -0.733 & -0.865 & -0.525 & -1.049 \\
\hline
\end{tabular}
}
\end{center}
\end{table}

\begin{table*}
\caption{{\bf Group Luminosity Function properties:} The bright-to-control
  ($\BC$) and bright-to-faint ($\BF$) ratios of the passive and
  star forming galaxy populations in the CNOC2 groups and the
  field. Additional collumns give the enhancement of $\BC$ or $\BF$ in
  the groups relative to the field with associated error ($\dBC \pm
  \sigma(\dBC)$, $\dBF \pm \sigma(\dBF)$) and significance of each
  enhancement ($\SBC$, $\SBF$).} 
\label{tab:BCBF}
\begin{center}
\begin{footnotesize}
\begin{tabular}{ccccccccccc}
\hline\hline
$\svi$(km~s$^{-1}$) & p/sf & $\BC$ & $\dBC$ & $\sigma(\dBC)$ & $\SBC$ & $\BF$ & $\dBF$ &$\sigma(\dBF)$ & $\SBF$\\
\hline
All & All & -0.425 & 0.18 & 0.12 & 85.6\% & -0.971 & 0.43 & 0.13 & 99.0\%\\
All & p & -0.29 & 0.12 & 0.16 & 53.1\% & -0.52 & 0.21 & 0.18 & 79.1\%\\
All & sf & -0.64 & 0.10 & 0.19 & 40.4\% & -1.14 & 0.30 & 0.19 & 88.8\%\\
0-400 & p & -0.434 & -0.03 & 0.19 & 13.6\% & -0.593 & 0.19 & 0.22 & 64.9\%\\
0-400 & sf & -0.453 & 0.28 & 0.22 & 80.3\% & -1.269 & 0.69 & 0.23 & 99.7\%\\
400-600 & p & -0.118 & 0.28 & 0.23 & 78.3\% & -0.459 & 0.65 & 0.26 & 97.7\%\\
400-600 & sf & -1.098 & -0.35 & 0.29 & 72.9\% & -0.963 & -0.43 & 0.31 & 78.3\%\\
600-1000 & p & 0.129 & 0.53 & 0.26 & 95.4\% & -0.424 & 0.08 & 0.30 & 20.2\%\\
600-1000 & sf & -1.159 & -0.41 & 0.26 & 91.8\% & -0.971 & -0.12 & 0.39 & 30.6\%\\
\hline
\end{tabular}
\end{footnotesize}
\end{center}
\end{table*}

\section{Discussion and Conclusions}

We have used the LDSS2 spectrograph on the Magellan telescope at LCO to
obtain 418 new galaxy redshifts in the regions of 26 CNOC2 groups at
intermediate redshift ($0.3 \leq z \leq 0.55$). Of these, 86 are group
members. With the depth of Magellan, we are highly complete and
unbiased down to $R_C = 22.0$ and out to 240\arcsec ($\sim 2r_{200}$
for a 500~km~s$^{-1}$ group at $z=0.3$). The primary purpose of this
paper is to present this data set. 

Our main result of this paper is the significant enhancement of $\fp$
in our group sample relative to the field
level, even in the smallest groups.  This indicates that star formation
is less prevalent among galaxies in the group environment than
in the field, even at intermediate redshift. 
This could be related to either a different formation history
in the group environment, intragroup environmental processes
accelerating galaxy evolution to the passive state, or a combination
of these two effects. 

We have investigated how this deficit of star forming galaxies depends
on the group properties, and find only relatively weak correlations.
Whilst there is little overall trend of $\fp$ with group velocity dispersion, 
there is a suggestion that $\fp$ is higher in the more concentrated systems with $\svi > 400$~km~s$^{-1}$.
We examined the radial dependence of $\fp$, finding a weak trend with radius. 
Comparing $\fp$ with the nearest neighbour distance, the passive fraction declines significantly at large separations, 
suggesting a relation to the substructure within groups.
Taking into account the low numbers of galaxies in each system, these trends are 
qualitatively similar to those reported in local group samples \citep[e.g.][]{Hashimoto98,Gomez03,Girardi03,Balogh03}. 
The correlations of passive fraction with environment also mimic the dependence of morphology on environment 
in local groups \citep[e.g.][]{PostmanGeller84,Zabludoff98,Zabludoff2000,Tran01}.

In rich clusters at the epoch of CNOC2, \citet{Fasano2000} find an evolving fraction of elliptical + S0 galaxy types with a fraction $f_{E+S0} \sim 0.45 \pm 0.15$ in clusters, within a clustercentric radius of $0.67h_{75}$Mpc and brighter than $M_V = -20$. This is comparible with the mean fraction of passive galaxies in the CNOC2 groups (down to brighter than \MBj$=-20.0$, see Figure~\ref{fig:fp_VD}) but with considerably less scatter in these richer systems. We note that the morphology-density relation is also effective in at least some of these clusters \citep[see also][]{Dressler97}, with a much higher fraction of early-type galaxies (approaching $f_{E+S0} = 1$) in the densest regions. However, the different selection (morphological, V-band limited) and our arbitrary division at EW[OII]=5\AA\ makes quantitative comparison difficult. The value of $\fp$ in groups is also significantly lower than seen in CNOC1 cluster cores at $z \sim 0.4$, with mean values of $\fp \sim 0.7$ within the inner $0.75h_{75}$~Mpc and brighter than \MBj$=-19.65$ \citep[][Note: whilst in that paper the division is at 10\AA, we use 5\AA\ for consistency]{Nakata04}. In Mulchaey et al., 2005 we will discuss the morphological properties of our sample in detail, using HST ACS imaging (now mostly complete).

We have coadded the data to provide a well understood stacked group
sample and a stacked field sample over the same redshift range ($0.3
\leq z \leq 0.55$), and use this sample to investigate the dependence
of $\fp$ on galaxy luminosity.  We find that $\fp$ is greatly enhanced
in group galaxies with respect to the field down to at least \MBj$\sim -19.0$.
In the brightest galaxies (\MBj$ \lsim -21.0$) this trend is strongest in the higher 
velocity dispersion systems ($\svi \gsim 400$~km~s$^{-1}$). 

We also investigate how the luminosity function of
galaxies depends on their environment: local studies have shown that
luminosity is at least as important as environment in determining
whether a galaxy is passive or star forming at the present day
\citep[e.g.][]{Blanton04}.
Our data suggest that the intermediate redshift
group galaxy population is enhanced in bright galaxies and that lower velocity
dispersion groups ($\svi \lsim 400$~km~s$^{-1}$) have a deficit of faint galaxies 
compared to the intermediate redshift field. However, only in the more massive groups is the enhanced bright 
population primarily passive ($\fp$ also enhanced).  

It is instructive to see how these trends compare with similar studies at low
redshift in the literature. The variation in the shape of the global
luminosity function measured from survey to survey is worryingly large
\citep{DriverDeProp03} and so it is important to use data from the
same survey where possible in studies of how the luminosity function
depends upon environment. Data from SDSS and 2dFGRS suggests that 
in regions of higher density the local galaxy population is biased towards a brighter 
characteristic magnitude and steeper faint-end slope \citep{DePropris03,Blanton03,Croton04}. 
This density dependence appears to be dominated by the relative abundance of early-type 
passive galaxies and may not be obvious in each individual cluster \citep[e.g. the Coma cluster,][]{Mobasher03}.
In local groups there may also be a correlation between group velocity dispersion 
and the faint-end slope, driven by the population of faint, early type galaxies 
\citep{Zabludoff2000,Christlein2000}. The large population of faint 
passive galaxies in our most massive groups ($\svi \gsim 600$~km~s$^{-1}$, see Figure~\ref{fig:lumfns_bysfandvd})
suggests that we might be seeing the same trends at $z \sim 0.45$, although there appears to be a 
deficit of faint galaxies in less massive groups, relative to the field.
The generation of this faint passive galaxy population is relavent to downsizing in 
the red sequence of cluster galaxies, with recent observations indicating that it does not 
extend faintwards of $\sim \Mstar + 2$ at $z \sim 1$ \citep{Kodama04,deLucia04}.

In Paper~II we will compare our CNOC2 groups to a low redshift group sample selected from the 
2dFGRS survey. In Paper~III we will examine the morphological composition of our groups, extending 
studies of the morphology-density relation to group environments at $z \sim 0.45$.

\section{Acknowledgements}\label{CNOC2Ack}

We would like to thank the staff of the Magellan telescope for their
tremendous support. RGB is supported by a PPARC Senior Research
Fellowship. DJW, MLB and RJW also thank PPARC for their support. We
are grateful to Dan Kelson for the use of his spectral reduction
software and to David Gilbank for his help when learning to use it. We
also acknowledge \emph{noao} for the wealth of {\sc iraf} tools used
during data reduction. We also acknowledge Tom Shanks and Phil Outram
for their observations at Magellan and the full CNOC2 team for the
outstanding dataset. Finally, we thank the anonymous referee for some 
useful feedback which has helped to improve this paper.


\begin{thebibliography}{}
\bibitem[{{Allington-Smith} {et~al.}(1993){Allington-Smith}, {Ellis}, {Zirbel},
  \& {Oemler}}]{AllSmith93}
{Allington-Smith}, J.~R., {Ellis}, R., {Zirbel}, E.~L., \& {Oemler}, A.~J.
  1993, ApJ, 404, 521

\bibitem[{{Baldry} {et~al.}(2004){Baldry}, {Glazebrook}, {Brinkmann}, {Ivezi{\'
  c}}, {Lupton}, {Nichol}, \& {Szalay}}]{Baldry03}
{Baldry}, I.~K., {Glazebrook}, K., {Brinkmann}, J., {Ivezi{\' c}}, {\v Z}.,
  {Lupton}, R.~H., {Nichol}, R.~C., \& {Szalay}, A.~S. 2004, ApJ, 600, 681

\bibitem[{{Balogh} {et~al.}(2004){Balogh}, {Eke}, {Miller}, {Lewis}, {Bower},
  {Couch}, {Nichol}, {Bland-Hawthorn}, {Baldry}, {Baugh}, {Bridges}, {Cannon},
  {Cole}, {Colless}, {Collins}, {Cross}, {Dalton}, {de Propris}, {Driver},
  {Efstathiou}, {Ellis}, {Frenk}, {Glazebrook}, {Gomez}, {Gray}, {Hawkins},
  {Jackson}, {Lahav}, {Lumsden}, {Maddox}, {Madgwick}, {Norberg}, {Peacock},
  {Percival}, {Peterson}, {Sutherland}, \& {Taylor}}]{Balogh03}
{Balogh}, M., {Eke}, V., {Miller}, C., {Lewis}, I., {Bower}, R., {Couch}, W.,
  {Nichol}, R., {Bland-Hawthorn}, J., {Baldry}, I.~K., {Baugh}, C., {Bridges},
  T., {Cannon}, R., {Cole}, S., {Colless}, M., {Collins}, C., {Cross}, N.,
  {Dalton}, G., {de Propris}, R., {Driver}, S.~P., {Efstathiou}, G., {Ellis},
  R.~S., {Frenk}, C.~S., {Glazebrook}, K., {Gomez}, P., {Gray}, A., {Hawkins},
  E., {Jackson}, C., {Lahav}, O., {Lumsden}, S., {Maddox}, S., {Madgwick}, D.,
  {Norberg}, P., {Peacock}, J.~A., {Percival}, W., {Peterson}, B.~A.,
  {Sutherland}, W., \& {Taylor}, K. 2004, MNRAS, 348, 1355

\bibitem[{{Beers} {et~al.}(1990){Beers}, {Flynn}, \& {Gebhardt}}]{Beers90}
{Beers}, T.~C., {Flynn}, K., \& {Gebhardt}, K. 1990, AJ, 100, 32

\bibitem[{{Blanton} {et~al.}(2004){Blanton}, {Eisenstein}, {Hogg}, {Schlegel},
  \& {Brinkmann}}]{Blanton04}
{Blanton}, M.~R., {Eisenstein}, D.~J., {Hogg}, D.~W., {Schlegel}, D.~J., \&
  {Brinkmann}, J. 2004, ApJ, submitted, astro-ph/0310453

\bibitem[{{Blanton} {et~al.}(2003){Blanton}, {Hogg}, {Bahcall}, {Baldry},
  {Brinkmann}, {Csabai}, {Eisenstein}, {Fukugita}, {Gunn}, {Ivezi{\' c}},
  {Lamb}, {Lupton}, {Loveday}, {Munn}, {Nichol}, {Okamura}, {Schlegel},
  {Shimasaku}, {Strauss}, {Vogeley}, \& {Weinberg}}]{Blanton03}
{Blanton}, M.~R., {Hogg}, D.~W., {Bahcall}, N.~A., {Baldry}, I.~K.,
  {Brinkmann}, J., {Csabai}, I., {Eisenstein}, D., {Fukugita}, M., {Gunn},
  J.~E., {Ivezi{\' c}}, {\v Z}., {Lamb}, D.~Q., {Lupton}, R.~H., {Loveday}, J.,
  {Munn}, J.~A., {Nichol}, R.~C., {Okamura}, S., {Schlegel}, D.~J.,
  {Shimasaku}, K., {Strauss}, M.~A., {Vogeley}, M.~S., \& {Weinberg}, D.~H.
  2003, ApJ, 594, 186

\bibitem[{{Bower} \& {Balogh}(2004)}]{BowerReview04}
{Bower}, R.~G. \& {Balogh}, M.~L. 2004, in Clusters of Galaxies: Probes of
  Cosmological Structure and Galaxy Evolution, 326--+

\bibitem[{{Brinchmann} {et~al.}(2004){Brinchmann}, {Charlot}, {White},
  {Tremonti}, {Kauffmann}, {Heckman}, \& {Brinkmann}}]{Brinchmann03}
{Brinchmann}, J., {Charlot}, S., {White}, S.~D.~M., {Tremonti}, C.,
  {Kauffmann}, G., {Heckman}, T., \& {Brinkmann}, J. 2004, MNRAS, 351, 1151

\bibitem[{Butcher \& Oemler(1984)}]{BO84}
Butcher, H. \& Oemler, A. 1984, ApJ, 285, 426

\bibitem[{{Carlberg} {et~al.}(2001){Carlberg}, {Yee}, {Morris}, {Lin}, {Hall},
  {Patton}, {Sawicki}, \& {Shepherd}}]{Carlberg01}
{Carlberg}, R.~G., {Yee}, H.~K.~C., {Morris}, S.~L., {Lin}, H., {Hall}, P.~B.,
  {Patton}, D.~R., {Sawicki}, M., \& {Shepherd}, C.~W. 2001, ApJ, 552, 427

\bibitem[{{Christlein}(2000)}]{Christlein2000}
{Christlein}, D. 2000, ApJ, 545, 145

\bibitem[{{Croton} {et~al.}(2004){Croton}, {Farrar}, {Norberg}, {Colless},
  {Peacock}, {Baldry}, {Baugh}, {Bland-Hawthorn}, {Bridges}, {Cannon}, {Cole},
  {Collins}, {Couch}, {Dalton}, {De Propris}, {Driver}, {Efstathiou}, {Ellis},
  {Frenk}, {Glazebrook}, {Jackson}, {Lahav}, {Lewis}, {Lumsden}, {Maddox},
  {Madgwick}, {Peterson}, {Sutherland}, \& {Taylor}}]{Croton04}
{Croton}, D.~J., {Farrar}, G.~R., {Norberg}, P., {Colless}, M., {Peacock},
  J.~A., {Baldry}, I.~K., {Baugh}, C.~M., {Bland-Hawthorn}, J., {Bridges}, T.,
  {Cannon}, R., {Cole}, S., {Collins}, C., {Couch}, W., {Dalton}, G., {De
  Propris}, R., {Driver}, S.~P., {Efstathiou}, G., {Ellis}, R.~S., {Frenk},
  C.~S., {Glazebrook}, K., {Jackson}, C., {Lahav}, O., {Lewis}, I., {Lumsden},
  S., {Maddox}, S., {Madgwick}, D., {Peterson}, B.~A., {Sutherland}, W., \&
  {Taylor}, K. 2004, astro-ph/, in press:, astro

\bibitem[{{De Lucia} {et~al.}(2004){De Lucia}, {Poggianti}, {Arag{\'
  o}n-Salamanca}, {Clowe}, {Halliday}, {Jablonka}, {Milvang-Jensen}, {Pell{\'
  o}}, {Poirier}, {Rudnick}, {Saglia}, {Simard}, \& {White}}]{deLucia04}
{De Lucia}, G., {Poggianti}, B.~M., {Arag{\' o}n-Salamanca}, A., {Clowe}, D.,
  {Halliday}, C., {Jablonka}, P., {Milvang-Jensen}, B., {Pell{\' o}}, R.,
  {Poirier}, S., {Rudnick}, G., {Saglia}, R., {Simard}, L., \& {White},
  S.~D.~M. 2004, ApJL, 610, L77

\bibitem[{{De Propris} {et~al.}(2003){De Propris}, {Colless}, {Driver},
  {Couch}, {Peacock}, {Baldry}, {Baugh}, {Bland-Hawthorn}, {Bridges}, {Cannon},
  {Cole}, {Collins}, {Cross}, {Dalton}, {Efstathiou}, {Ellis}, {Frenk},
  {Glazebrook}, {Hawkins}, {Jackson}, {Lahav}, {Lewis}, {Lumsden}, {Maddox},
  {Madgwick}, {Norberg}, {Percival}, {Peterson}, {Sutherland}, \&
  {Taylor}}]{DePropris03}
{De Propris}, R., {Colless}, M., {Driver}, S.~P., {Couch}, W., {Peacock},
  J.~A., {Baldry}, I.~K., {Baugh}, C.~M., {Bland-Hawthorn}, J., {Bridges}, T.,
  {Cannon}, R., {Cole}, S., {Collins}, C., {Cross}, N., {Dalton}, G.~B.,
  {Efstathiou}, G., {Ellis}, R.~S., {Frenk}, C.~S., {Glazebrook}, K.,
  {Hawkins}, E., {Jackson}, C., {Lahav}, O., {Lewis}, I., {Lumsden}, S.,
  {Maddox}, S., {Madgwick}, D.~S., {Norberg}, P., {Percival}, W., {Peterson},
  B., {Sutherland}, W., \& {Taylor}, K. 2003, MNRAS, 342, 725

\bibitem[{Dressler(1980)}]{Dres80disc}
Dressler, A. 1980, ApJ, 236, 351

\bibitem[{{Dressler} {et~al.}(1997){Dressler}, {Oemler}, {Couch}, {Smail},
  {Ellis}, {Barger}, {Butcher}, {Poggianti}, \& {Sharples}}]{Dressler97}
{Dressler}, A., {Oemler}, A.~J., {Couch}, W.~J., {Smail}, I., {Ellis}, R.~S.,
  {Barger}, A., {Butcher}, H., {Poggianti}, B.~M., \& {Sharples}, R.~M. 1997,
  ApJ, 490, 577

\bibitem[{{Dressler} \& {Shectman}(1987)}]{DressSchect87}
{Dressler}, A. \& {Shectman}, S.~A. 1987, AJ, 94, 899

\bibitem[{{Driver} \& {De Propris}(2003)}]{DriverDeProp03}
{Driver}, S. \& {De Propris}, R. 2003, Astrophysics and Space Science, 285, 175

\bibitem[{{Efron}(1982)}]{Jackknife}
{Efron}, B. 1982, {The Jackknife, the Bootstrap and other resampling plans}
  (CBMS-NSF Regional Conference Series in Applied Mathematics, Philadelphia:
  Society for Industrial and Applied Mathematics (SIAM), 1982)

\bibitem[{{Eke} {et~al.}(2004){Eke}, {Baugh}, {Cole}, {Frenk}, {Norberg},
  {Peacock}, {Baldry}, {Bland-Hawthorn}, {Bridges}, {Cannon}, {Colless},
  {Collins}, {Couch}, {Dalton}, {de Propris}, {Driver}, {Efstathiou}, {Ellis},
  {Glazebrook}, {Jackson}, {Lahav}, {Lewis}, {Lumsden}, {Maddox}, {Madgwick},
  {Peterson}, {Sutherland}, \& {Taylor}}]{Eke04}
{Eke}, V.~R., {Baugh}, C.~M., {Cole}, S., {Frenk}, C.~S., {Norberg}, P.,
  {Peacock}, J.~A., {Baldry}, I.~K., {Bland-Hawthorn}, J., {Bridges}, T.,
  {Cannon}, R., {Colless}, M., {Collins}, C., {Couch}, W., {Dalton}, G., {de
  Propris}, R., {Driver}, S.~P., {Efstathiou}, G., {Ellis}, R.~S.,
  {Glazebrook}, K., {Jackson}, C., {Lahav}, O., {Lewis}, I., {Lumsden}, S.,
  {Maddox}, S., {Madgwick}, D., {Peterson}, B.~A., {Sutherland}, W., \&
  {Taylor}, K. 2004, MNRAS, 348, 866

\bibitem[{{Fasano} {et~al.}(2000){Fasano}, {Poggianti}, {Couch}, {Bettoni},
  {Kj{\ae}rgaard}, \& {Moles}}]{Fasano2000}
{Fasano}, G., {Poggianti}, B.~M., {Couch}, W.~J., {Bettoni}, D.,
  {Kj{\ae}rgaard}, P., \& {Moles}, M. 2000, ApJ, 542, 673

\bibitem[{{Flores} {et~al.}(2004){Flores}, {Hammer}, {Elbaz}, {Cesarsky},
  {Liang}, {Fadda}, \& {Gruel}}]{Flores03}
{Flores}, H., {Hammer}, F., {Elbaz}, D., {Cesarsky}, C.~J., {Liang}, Y.~C.,
  {Fadda}, D., \& {Gruel}, N. 2004, A\&A, 415, 885

\bibitem[{{G{\' o}mez} {et~al.}(2003){G{\' o}mez}, {Nichol}, {Miller},
  {Balogh}, {Goto}, {Zabludoff}, {Romer}, {Bernardi}, {Sheth}, {Hopkins},
  {Castander}, {Connolly}, {Schneider}, {Brinkmann}, {Lamb}, {SubbaRao}, \&
  {York}}]{Gomez03}
{G{\' o}mez}, P.~L., {Nichol}, R.~C., {Miller}, C.~J., {Balogh}, M.~L., {Goto},
  T., {Zabludoff}, A.~I., {Romer}, A.~K., {Bernardi}, M., {Sheth}, R.,
  {Hopkins}, A.~M., {Castander}, F.~J., {Connolly}, A.~J., {Schneider}, D.~P.,
  {Brinkmann}, J., {Lamb}, D.~Q., {SubbaRao}, M., \& {York}, D.~G. 2003, ApJ,
  584, 210

\bibitem[{{Girardi} {et~al.}(2003){Girardi}, {Rigoni}, {Mardirossian}, \&
  {Mezzetti}}]{Girardi03}
{Girardi}, M., {Rigoni}, E., {Mardirossian}, F., \& {Mezzetti}, M. 2003, A\&A,
  406, 403

\bibitem[{{Hashimoto} {et~al.}(1998){Hashimoto}, {Oemler}, {Lin}, \&
  {Tucker}}]{Hashimoto98}
{Hashimoto}, Y., {Oemler}, A.~J., {Lin}, H., \& {Tucker}, D.~L. 1998, ApJ, 499,
  589

\bibitem[{{Hickson} {et~al.}(1989){Hickson}, {Kindl}, \& {Auman}}]{Hickson89}
{Hickson}, P., {Kindl}, E., \& {Auman}, J.~R. 1989, ApJS, 70, 687

\bibitem[{{Hopkins}(2004)}]{Hopkins04}
{Hopkins}, A.~M. 2004, ArXiv Astrophysics e-prints

\bibitem[{{Hopkins} {et~al.}(2003){Hopkins}, {Miller}, {Nichol}, {Connolly},
  {Bernardi}, {G{\' o}mez}, {Goto}, {Tremonti}, {Brinkmann}, {Ivezi{\' c}}, \&
  {Lamb}}]{Hopkins03}
{Hopkins}, A.~M., {Miller}, C.~J., {Nichol}, R.~C., {Connolly}, A.~J.,
  {Bernardi}, M., {G{\' o}mez}, P.~L., {Goto}, T., {Tremonti}, C.~A.,
  {Brinkmann}, J., {Ivezi{\' c}}, {\v Z}., \& {Lamb}, D.~Q. 2003, ApJ, 599, 971

\bibitem[{{Huchra} \& {Geller}(1982)}]{HuchraGeller82}
{Huchra}, J.~P. \& {Geller}, M.~J. 1982, ApJ, 257, 423

\bibitem[{{Jansen} {et~al.}(2001){Jansen}, {Franx}, \& {Fabricant}}]{Jansen01}
{Jansen}, R.~A., {Franx}, M., \& {Fabricant}, D. 2001, ApJ, 551, 825

\bibitem[{{Jones} {et~al.}(2002){Jones}, {McHardy}, {Newsam}, \&
  {Mason}}]{Jones02b}
{Jones}, L.~R., {McHardy}, I., {Newsam}, A., \& {Mason}, K. 2002, MNRAS, 334,
  219

\bibitem[{{Kauffmann} {et~al.}(2004){Kauffmann}, {White}, {Heckman}, {M{\'
  e}nard}, {Brinchmann}, {Charlot}, {Tremonti}, \& {Brinkmann}}]{Kauffmann04}
{Kauffmann}, G., {White}, S.~D.~M., {Heckman}, T.~M., {M{\' e}nard}, B.,
  {Brinchmann}, J., {Charlot}, S., {Tremonti}, C., \& {Brinkmann}, J. 2004,
  MNRAS, 314

\bibitem[{{Kelson}(2003)}]{Kelson03}
{Kelson}, D.~D. 2003, PASP, 115, 688

\bibitem[{{Kennicutt}(1992)}]{Kennicutt92}
{Kennicutt}, R.~C. 1992, ApJS, 79, 255

\bibitem[{{Kodama} {et~al.}(2001){Kodama}, {Smail}, {Nakata}, {Okamura}, \&
  {Bower}}]{Kodama01}
{Kodama}, T., {Smail}, I., {Nakata}, F., {Okamura}, S., \& {Bower}, R.~G. 2001,
  ApJL, 562, L9

\bibitem[{{Kodama} {et~al.}(2004){Kodama}, {Yamada}, {Akiyama}, {Aoki}, {Doi},
  {Furusawa}, {Fuse}, {Imanishi}, {Ishida}, {Iye}, {Kajisawa}, {Karoji},
  {Kobayashi}, {Komiyama}, {Kosugi}, {Maeda}, {Miyazaki}, {Mizumoto},
  {Morokuma}, {Nakata}, {Noumaru}, {Ogasawara}, {Ouchi}, {Sasaki}, {Sekiguchi},
  {Shimasaku}, {Simpson}, {Takata}, {Tanaka}, {Ueda}, {Yasuda}, \&
  {Yoshida}}]{Kodama04}
{Kodama}, T., {Yamada}, T., {Akiyama}, M., {Aoki}, K., {Doi}, M., {Furusawa},
  H., {Fuse}, T., {Imanishi}, M., {Ishida}, C., {Iye}, M., {Kajisawa}, M.,
  {Karoji}, H., {Kobayashi}, N., {Komiyama}, Y., {Kosugi}, G., {Maeda}, Y.,
  {Miyazaki}, S., {Mizumoto}, Y., {Morokuma}, T., {Nakata}, F., {Noumaru}, J.,
  {Ogasawara}, R., {Ouchi}, M., {Sasaki}, T., {Sekiguchi}, K., {Shimasaku}, K.,
  {Simpson}, C., {Takata}, T., {Tanaka}, I., {Ueda}, Y., {Yasuda}, N., \&
  {Yoshida}, M. 2004, MNRAS, 350, 1005

\bibitem[{{Kurtz} \& {Mink}(1998)}]{rvsao98}
{Kurtz}, M.~J. \& {Mink}, D.~J. 1998, PASP, 110, 934

\bibitem[{{Lilly} {et~al.}(1996){Lilly}, {Le Fevre}, {Hammer}, \&
  {Crampton}}]{Lilly96}
{Lilly}, S.~J., {Le Fevre}, O., {Hammer}, F., \& {Crampton}, D. 1996, ApJL,
  460, L1+

\bibitem[{{Lin} {et~al.}(1999){Lin}, {Yee}, {Carlberg}, {Morris}, {Sawicki},
  {Patton}, {Wirth}, \& {Shepherd}}]{Lin99}
{Lin}, H., {Yee}, H.~K.~C., {Carlberg}, R.~G., {Morris}, S.~L., {Sawicki}, M.,
  {Patton}, D.~R., {Wirth}, G., \& {Shepherd}, C.~W. 1999, ApJ, 518, 533

\bibitem[{Madau {et~al.}(1998)Madau, Pozzetti, \& Dickinson}]{Madau98}
Madau, P., Pozzetti, L., \& Dickinson, M. 1998, ApJ, 498, 106

\bibitem[{{Mart{\'{\i}}nez} {et~al.}(2002){Mart{\'{\i}}nez}, {Zandivarez},
  {Dom{\'{\i}}nguez}, {Merch{\' a}n}, \& {Lambas}}]{Martinez02}
{Mart{\'{\i}}nez}, H.~J., {Zandivarez}, A., {Dom{\'{\i}}nguez}, M., {Merch{\'
  a}n}, M.~E., \& {Lambas}, D.~G. 2002, MNRAS, 333, L31

\bibitem[{{Mobasher} {et~al.}(2003){Mobasher}, {Colless}, {Carter},
  {Poggianti}, {Bridges}, {Kranz}, {Komiyama}, {Kashikawa}, {Yagi}, \&
  {Okamura}}]{Mobasher03}
{Mobasher}, B., {Colless}, M., {Carter}, D., {Poggianti}, B.~M., {Bridges},
  T.~J., {Kranz}, K., {Komiyama}, Y., {Kashikawa}, N., {Yagi}, M., \&
  {Okamura}, S. 2003, ApJ, 587, 605

\bibitem[{{Mulchaey} {et~al.}(2003){Mulchaey}, {Davis}, {Mushotzky}, \&
  {Burstein}}]{Mulchaey03}
{Mulchaey}, J.~S., {Davis}, D.~S., {Mushotzky}, R.~F., \& {Burstein}, D. 2003,
  ApJS, 145, 39

\bibitem[{{Nakata} {et~al.}(2004){Nakata}, G., {Balogh}, \& J.}]{Nakata04}
{Nakata}, F., G., B.~R., {Balogh}, M.~L., \& J., W.~D. 2004, MNRAS, accepted

\bibitem[{{Norberg} {et~al.}(2002){Norberg}, {Cole}, {Baugh}, {Frenk},
  {Baldry}, {Bland-Hawthorn}, {Bridges}, {Cannon}, {Colless}, {Collins},
  {Couch}, {Cross}, {Dalton}, {De Propris}, {Driver}, {Efstathiou}, {Ellis},
  {Glazebrook}, {Jackson}, {Lahav}, {Lewis}, {Lumsden}, {Maddox}, {Madgwick},
  {Peacock}, {Peterson}, {Sutherland}, \& {Taylor}}]{Norberg02}
{Norberg}, P., {Cole}, S., {Baugh}, C.~M., {Frenk}, C.~S., {Baldry}, I.,
  {Bland-Hawthorn}, J., {Bridges}, T., {Cannon}, R., {Colless}, M., {Collins},
  C., {Couch}, W., {Cross}, N.~J.~G., {Dalton}, G., {De Propris}, R., {Driver},
  S.~P., {Efstathiou}, G., {Ellis}, R.~S., {Glazebrook}, K., {Jackson}, C.,
  {Lahav}, O., {Lewis}, I., {Lumsden}, S., {Maddox}, S., {Madgwick}, D.,
  {Peacock}, J.~A., {Peterson}, B.~A., {Sutherland}, W., \& {Taylor}, K. 2002,
  MNRAS, 336, 907

\bibitem[{{Poggianti} {et~al.}(1999){Poggianti}, {Smail}, {Dressler}, {Couch},
  {Barger}, {Butcher}, {Ellis}, \& {Oemler}}]{Poggianti99}
{Poggianti}, B.~M., {Smail}, I., {Dressler}, A., {Couch}, W.~J., {Barger},
  A.~J., {Butcher}, H., {Ellis}, R.~S., \& {Oemler}, A.~J. 1999, ApJ, 518, 576

\bibitem[{{Postman} \& {Geller}(1984)}]{PostmanGeller84}
{Postman}, M. \& {Geller}, M.~J. 1984, ApJ, 281, 95

\bibitem[{{Ramella} {et~al.}(1989){Ramella}, {Geller}, \& {Huchra}}]{Ramella89}
{Ramella}, M., {Geller}, M.~J., \& {Huchra}, J.~P. 1989, ApJ, 344, 57

\bibitem[{{Ramella} {et~al.}(1997){Ramella}, {Pisani}, \& {Geller}}]{Ramella97}
{Ramella}, M., {Pisani}, A., \& {Geller}, M.~J. 1997, AJ, 113, 483

\bibitem[{{Ramella} {et~al.}(1999){Ramella}, {Zamorani}, {Zucca}, {Stirpe},
  {Vettolani}, {Balkowski}, {Blanchard}, {Cappi}, {Cayatte}, {Chincarini},
  {Collins}, {Guzzo}, {MacGillivray}, {Maccagni}, {Maurogordato}, {Merighi},
  {Mignoli}, {Pisani}, {Proust}, \& {Scaramella}}]{Ramella99}
{Ramella}, M., {Zamorani}, G., {Zucca}, E., {Stirpe}, G.~M., {Vettolani}, G.,
  {Balkowski}, C., {Blanchard}, A., {Cappi}, A., {Cayatte}, V., {Chincarini},
  G., {Collins}, C., {Guzzo}, L., {MacGillivray}, H., {Maccagni}, D.,
  {Maurogordato}, S., {Merighi}, R., {Mignoli}, M., {Pisani}, A., {Proust}, D.,
  \& {Scaramella}, R. 1999, A\&A, 342, 1

\bibitem[{{Schlegel} {et~al.}(1998){Schlegel}, {Finkbeiner}, \&
  {Davis}}]{Schlegel98}
{Schlegel}, D.~J., {Finkbeiner}, D.~P., \& {Davis}, M. 1998, apj, 500, 525

\bibitem[{{Severgnini} \& {Saracco}(2001)}]{SS01}
{Severgnini}, P. \& {Saracco}, P. 2001, Astrophysics and Space Science, 276,
  749

\bibitem[{{Shepherd} {et~al.}(2001){Shepherd}, {Carlberg}, {Yee}, {Morris},
  {Lin}, {Sawicki}, {Hall}, \& {Patton}}]{Shepherd01}
{Shepherd}, C.~W., {Carlberg}, R.~G., {Yee}, H.~K.~C., {Morris}, S.~L., {Lin},
  H., {Sawicki}, M., {Hall}, P.~B., \& {Patton}, D.~R. 2001, ApJ, 560, 72

\bibitem[{{Strateva} {et~al.}(2001){Strateva}, {Ivezi{\' c}}, {Knapp},
  {Narayanan}, {Strauss}, {Gunn}, {Lupton}, {Schlegel}, {Bahcall}, {Brinkmann},
  {Brunner}, {Budav{\' a}ri}, {Csabai}, {Castander}, {Doi}, {Fukugita}, {Gy{\H
  o}ry}, {Hamabe}, {Hennessy}, {Ichikawa}, {Kunszt}, {Lamb}, {McKay},
  {Okamura}, {Racusin}, {Sekiguchi}, {Schneider}, {Shimasaku}, \&
  {York}}]{Strateva01}
{Strateva}, I., {Ivezi{\' c}}, {\v Z}., {Knapp}, G.~R., {Narayanan}, V.~K.,
  {Strauss}, M.~A., {Gunn}, J.~E., {Lupton}, R.~H., {Schlegel}, D., {Bahcall},
  N.~A., {Brinkmann}, J., {Brunner}, R.~J., {Budav{\' a}ri}, T., {Csabai}, I.,
  {Castander}, F.~J., {Doi}, M., {Fukugita}, M., {Gy{\H o}ry}, Z., {Hamabe},
  M., {Hennessy}, G., {Ichikawa}, T., {Kunszt}, P.~Z., {Lamb}, D.~Q., {McKay},
  T.~A., {Okamura}, S., {Racusin}, J., {Sekiguchi}, M., {Schneider}, D.~P.,
  {Shimasaku}, K., \& {York}, D. 2001, AJ, 122, 1861

\bibitem[{{Tonry} \& {Davis}(1979)}]{TD79}
{Tonry}, J. \& {Davis}, M. 1979, AJ, 84, 1511

\bibitem[{{Tran} {et~al.}(2001){Tran}, {Simard}, {Zabludoff}, \&
  {Mulchaey}}]{Tran01}
{Tran}, K.~H., {Simard}, L., {Zabludoff}, A.~I., \& {Mulchaey}, J.~S. 2001,
  ApJ, 549, 172

\bibitem[{{Tucker} {et~al.}(2000){Tucker}, {Oemler}, {Hashimoto}, {Shectman},
  {Kirshner}, {Lin}, {Landy}, {Schechter}, \& {Allam}}]{Tucker2000}
{Tucker}, D.~L., {Oemler}, A.~J., {Hashimoto}, Y., {Shectman}, S.~A.,
  {Kirshner}, R.~P., {Lin}, H., {Landy}, S.~D., {Schechter}, P.~L., \& {Allam},
  S.~S. 2000, ApJS, 130, 237

\bibitem[{{Whitaker} {et~al.}(2004){Whitaker}, {Morris}, \& {The CNOC2
  Team}}]{Whitaker04}
{Whitaker}, R.~J., {Morris}, S.~L., \& {The CNOC2 Team}. 2004, MNRAS, in
  preparation

\bibitem[{{Wilman} {et~al.}(2004){Wilman}, {Balogh}, {Bower}, {Mulchaey},
  {Oemler Jr}, {Carlberg}, {Eke}, {Lewis}, {Morris}, \& {Whitaker}}]{Wilman04b}
{Wilman}, D.~J., {Balogh}, M.~L., {Bower}, R.~G., {Mulchaey}, J.~S., {Oemler
  Jr}, A., {Carlberg}, R.~G., {Eke}, V.~R., {Lewis}, I.~J., {Morris}, S.~L., \&
  {Whitaker}, R.~J. 2004, MNRAS, accepted

\bibitem[{{Wilson} {et~al.}(2002){Wilson}, {Cowie}, {Barger}, \&
  {Burke}}]{Wilson02}
{Wilson}, G., {Cowie}, L.~L., {Barger}, A.~J., \& {Burke}, D.~J. 2002, AJ, 124,
  1258

\bibitem[{{Yee} {et~al.}(1996){Yee}, {Ellingson}, \& {Carlberg}}]{Yee96}
{Yee}, H.~K.~C., {Ellingson}, E., \& {Carlberg}, R.~G. 1996, ApJS, 102, 269

\bibitem[{{Yee} {et~al.}(2000){Yee}, {Morris}, {Lin}, {Carlberg}, {Hall},
  {Sawicki}, {Patton}, {Wirth}, {Ellingson}, \& {Shepherd}}]{Yee00}
{Yee}, H.~K.~C., {Morris}, S.~L., {Lin}, H., {Carlberg}, R.~G., {Hall}, P.~B.,
  {Sawicki}, M., {Patton}, D.~R., {Wirth}, G.~D., {Ellingson}, E., \&
  {Shepherd}, C.~W. 2000, ApJS, 129, 475

\bibitem[{{Zabludoff} \& {Mulchaey}(1998)}]{Zabludoff98}
{Zabludoff}, A.~I. \& {Mulchaey}, J.~S. 1998, ApJ, 496, 39

\bibitem[{{Zabludoff} \& {Mulchaey}(2000)}]{Zabludoff2000}
---. 2000, ApJ, 539, 136

\end{thebibliography}

\appendix
\section{Completeness in the stacked group}\label{sec:completeness}

Despite the greater depth and spectroscopic completeness achieved in the region of the groups with our Magellan targetting, it was still necessary to adopt a sparse sampling strategy due to the high density of targets, especially close to our new magnitude limit of $R_{c} = 22$. In this section, we investigate the selection functions for the spectroscopic sample, seeking a representative means of stacking the data. The CNOC2 photometric catalogue is complete to $R_{c}\sim 23$ and so the probability that we possess a redshift of any galaxy brighter than this limit can be easily understood. In Section~\ref{sec:seleffects} we assume that our Magellan data reach sufficient depth to be unbiased at magnitudes $R_{c} \leq 22$. We then describe a simple weighting scheme which accounts for targetting bias in both CNOC2 and Magellan spectroscopic samples. This scheme differentiates between galaxies with redshifts from each source which have different selection functions. In Section~\ref{sec:zincompl} we test our assumption of unbiased redshift completeness in the Magellan data and show that this assumption holds within the limits of tests using current data.
 
\subsection{Selection Functions}\label{sec:seleffects}
\begin{figure}
\centerline{\psfig{figure=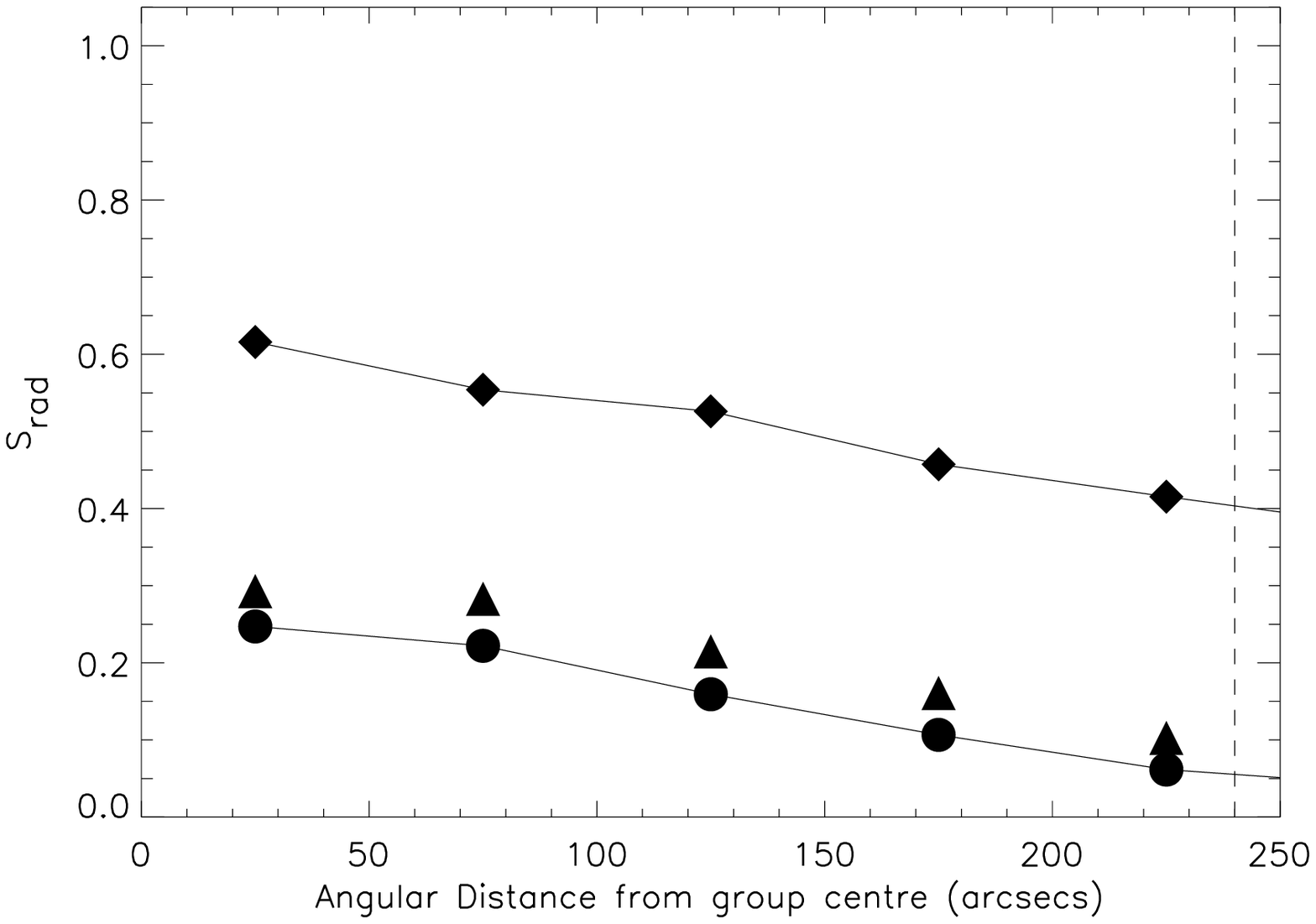,width=0.5\textwidth}}
\vskip .3cm
\caption{Radial selection function of the stacked data as a function of
  angular distance from the target group centre. Filled circles
  represent the fraction of galaxies with Magellan redshifts
  ($S_{rad}(L_{z})$); triangles represent the fraction of galaxies
  targetted by Magellan ($S_{rad}(L_{obs})$) and diamonds represent
  galaxies with CNOC2 or Magellan redshifts ($S_{rad}(L_{z}) +
  S_{rad}(C_{z})$). Only galaxies within 240\arcsec\ and with $R_C \leq
  22.0$ are considered in our analysis.} 
\label{fig:radcompl}
\end{figure}
To understand the selection functions for the spectroscopic sample, it
must first be split into its component CNOC2 and Magellan subsets as
selection depends critically on the observing strategies and facilities
used. Target selection in the CNOC2 spectroscopic survey was primarily
dependent upon apparent $R_{c}$-band magnitude. However, below the
nominal spectroscopic limit of $R_{c}=21.5$ in that survey, galaxies
with redshifts become biased towards strong EW[OII] with respect to
those obtained using Magellan. In contrast target selection with
Magellan is dependent upon both magnitude and angular distance from the
targetted group centre. However there appears to be no strong bias
towards emission redshifts brighter than our magnitude limit of
$R_{c}=22$ (see Section~\ref{sec:zincompl}). Therefore we can implement
a simple weighting scheme which simulates a $100$ per cent  complete
survey and removes targetting bias and CNOC2 redshift incompleteness
bias simultaneously: 

The weighting scheme requires a radial selection function and a
magnitude-dependent selection function. For each galaxy in the
spectroscopic sample, a radial weight $W_{rad}$ and a
magnitude-dependent weight, $W_{mag}$ are computed in such a way that
the total weight of all galaxies with redshifts (measured-z galaxies)
$\Sigma_{1}^{N_{spec}} (W_{rad}.W_{mag}) = N_{phot}$, the total number
of galaxies in the photometric sample brighter than $R_{c}=22$. We
denote galaxies with redshifts from the original CNOC2 survey with the
suffix $(C_{z})$ and those with Magellan LDSS2 redshifts with the
suffix $(L_{z})$. 

We begin by looking at the radial selection function, $S_{rad}$. The
fraction of galaxies with CNOC2 redshifts shows no significant
dependence upon angular distance from the centre of the nearest group
and so we choose $W_{rad}(C_{z}) = 1$ for all these galaxies. However
there is a strong dependence on angular radius for the fraction of
Magellan redshifts (or Magellan targets). In Figure~\ref{fig:radcompl},
the fraction of galaxies in the CNOC2 photometric catalogue ($R_{c}
\leq 22$) targetted by Magellan, $S_{rad}(L_{obs})$ (triangles), where
redshifts were obtained, $S_{rad}(L_{z})$ (filled circles) and
including the fraction with CNOC2 redshifts, $S_{rad}(L_{z}) +
S_{rad}(C_{z})$ (diamonds) is shown as a function of angular
radius. Most untargetted galaxies lie in the magnitude range $21.5 \leq
R_{c} \leq 22$. We apply a final cut to stacked data at 240\arcsec\,
which is the approximate limiting radius for LDSS2 targetting. This
corresponds to $1h_{75}^{-1}$Mpc at the low redshift end of our sample,
$z=0.3$. The linear spline through the filled circles
($S_{rad}(L_{z})$) represents the radial selection function
$S_{rad}(L_{z})$ from which the value of $S_{rad}$ is linearly
interpolated for any galaxy targetted with Magellan. For these
galaxies, the radial weight is then simply computed to be
$W_{rad}(L_{z}) = S_{0}(L_{z})/S_{rad}(L_{z})$, normalised so that
galaxies located at the group centre receive a weighting
$W_{rad}=1$. We note that a galaxy at 240\arcsec\ receives a weighting
of $W_{rad}\sim 4.47$.  
\begin{figure}
\centerline{\psfig{figure=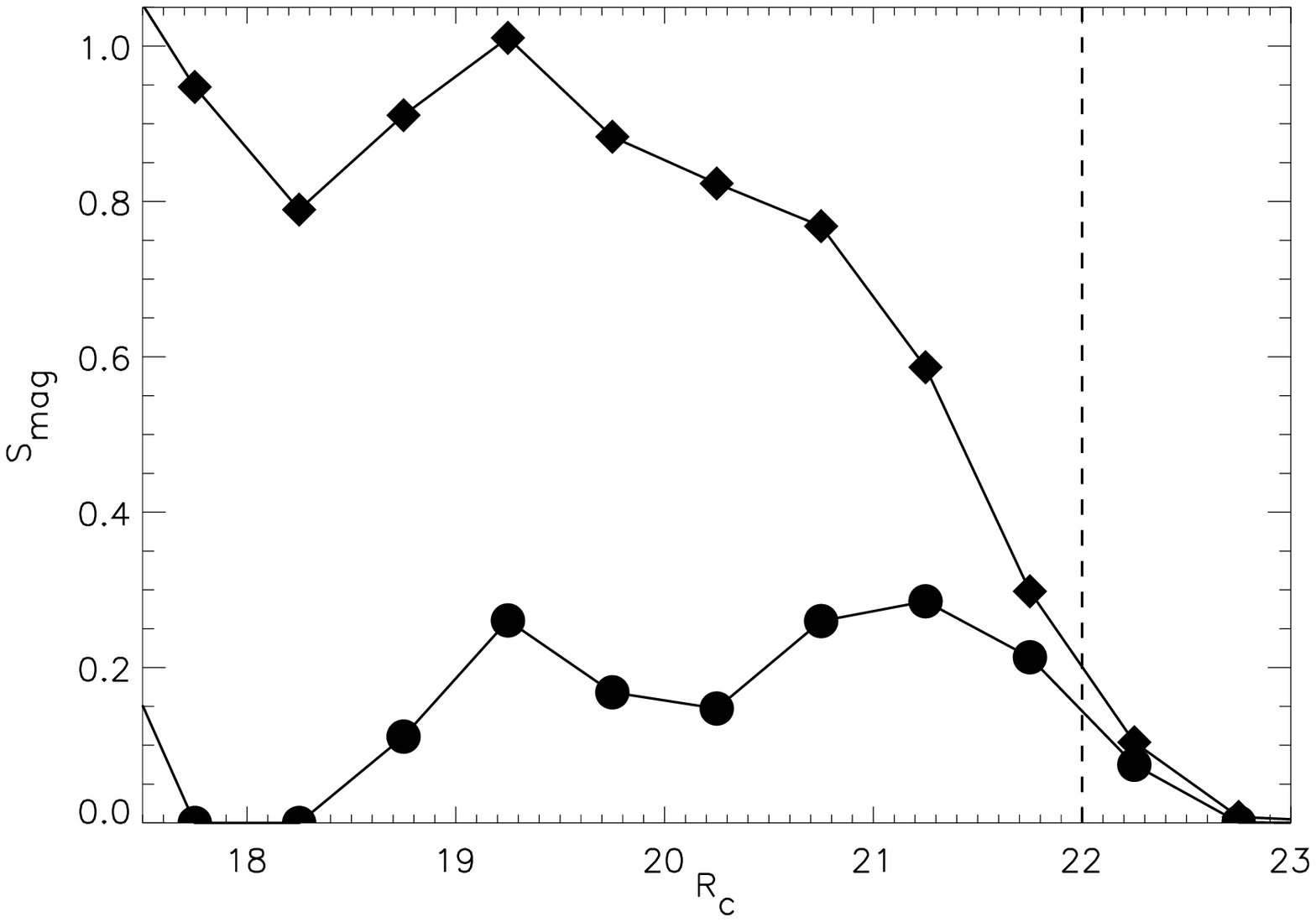,width=0.5\textwidth}}
\caption{Selection functions of the stacked group sample as a function
  of $R_{c}$-band magnitude. Filled circles represent the fraction of
  galaxies with Magellan redshifts, weighted by $W_{rad}$
  ($S_{mag}(L_{z})$) and diamonds represent galaxies with CNOC2 or
  Magellan redshifts ($S_{mag}(L_{z}) + S_{mag}(C_{z})$). Only galaxies
  brighter than $R_{c}=22$ are considered in our analaysis.} 
\label{fig:magcompl}
\end{figure}

In Figure~\ref{fig:magcompl}, we show the magnitude-dependent selection
function. Galaxies are limited to within 240\arcsec\ and galaxies with
redshifts are weighted by $W_{rad}$. As in Figure~\ref{fig:radcompl},
we plot the fraction of galaxies with Magellan redshifts,
$S_{mag}(L_{z})$ (filled circles) and the fraction with Magellan or
CNOC2 redshifts, $S_{mag}(L_{z}) + S_{mag}(C_{z})$
(diamonds). $W_{mag}$ must weight the galaxies such that
$\Sigma_{1}^{N_{spec}} (W_{rad}.W_{mag}) = N_{phot}$ in each magnitude
bin and the combined weight of all galaxies must accurately emulate the
properties of the entire galaxy population. In each bin of magnitude we
normalise the total weighted fraction of CNOC2 redshifts to be
equal to the fraction of CNOC2 redshifts in the unweighted
sample, $S_{mag}(C_{z})$. We know that in the entire population, that
fraction at least will have the properties of CNOC2 spectroscopic
galaxies, regardless of CNOC2 redshift incompleteness. Magellan targets
come from the remaining galaxies without CNOC2 redshifts (ignoring a
tiny comparison sample) and these have properties typical of that
population at any given magnitude (based upon our assumption of
unbiased redshift completeness in the Magellan data - see
Section~\ref{sec:zincompl}). Therefore the remainder of the combined
weight required to match the total number of galaxies in any magnitude
bin ($N_{phot}$) is spread amongst the galaxies with Magellan
redshifts. Thus $W_{mag}$ is computed independently for galaxies with
CNOC2 redshifts and Magellan redshifts using
equations~\ref{eqn:wmagczN} and~\ref{eqn:wmaglzN}, below (which simplify to
equations~\ref{eqn:wmagczS} and~\ref{eqn:wmaglzS}). We note that each
quantity is computed at a given $R_c$-band magnitude (or across
each magnitude bin). 

\begin{align}
W_{mag}(C_{z}) = 1 + \frac{\left[ N_{phot} - N_{C_{z}} - \Sigma (W_{rad}(L_z))\right]}{N_{phot}},
\label{eqn:wmagczN}
\end{align}

\noindent where $W_{mag}(L_{z}) = 1 + \alpha$ and
\begin{align}
\alpha = \frac{\left[N_{phot} - N_{C_{z}} - \Sigma (W_{rad}(L_z))\right]}{N_{phot}} \times \frac{N_{phot} - N_{C_{z}}}{\Sigma (W_{rad}(L_z))}.
\label{eqn:wmaglzN}
\end{align}
\newline
\noindent In terms of $S_{mag}(C_{z})$ and $S_{mag}(L_{z})$ these become:

\begin{align}
W_{mag}(C_{z}) = 1+ \left[ 1 - S_{mag}(C_{z}) - S_{mag}(L_{z}) \right]
\label{eqn:wmagczS}
\end{align}

\noindent and $\alpha$ in equation~\ref{eqn:wmaglzN} becomes:
\begin{align}
\alpha = \left[ 1 - S_{mag}(C_{z}) - S_{mag}(L_{z})\right] \times \frac{1 - S_{mag}(C_{z})}{S_{mag}(L_{z})},
\label{eqn:wmaglzS}
\end{align}

\noindent where $S_{mag}(C_{z})$ and $S_{mag}(L_{z})$ are computed by
linear interpolation through the points in
Figure~\ref{fig:magcompl}. The combined weight for each galaxy is
then given by 

\begin{equation}
W_C = W_{rad} \times W_{mag}.
\label{eqn:wfinal}
\end{equation}

We note from Figure~\ref{fig:magcompl} that in the magnitude bin at
$21.5 \leq R_c \leq 22$ the fraction of galaxies with redshifts is down
to $\sim 30$ per cent.  However, it is at magnitudes below $R_{c}=21$
that the Magellan data comes to dominate the statistics and the greater
depth of the larger telescope allows us to achieve unbiased and highly
successful redshift determination down to $R_{c}=22$. This fainter
galaxy population is expected to be more star forming than the brighter
population in the local Universe
\citep[e.g.][]{Brinchmann03}. Therefore by reaching this depth, we can
begin to probe the evolution of group galaxy properties since
intermediate redshift significantly below $\Mstar$ to reach this
interesting population of mainly star forming galaxies. 

\subsection{Magellan Redshift Completeness}\label{sec:zincompl}

\begin{figure}
\centerline{\psfig{figure=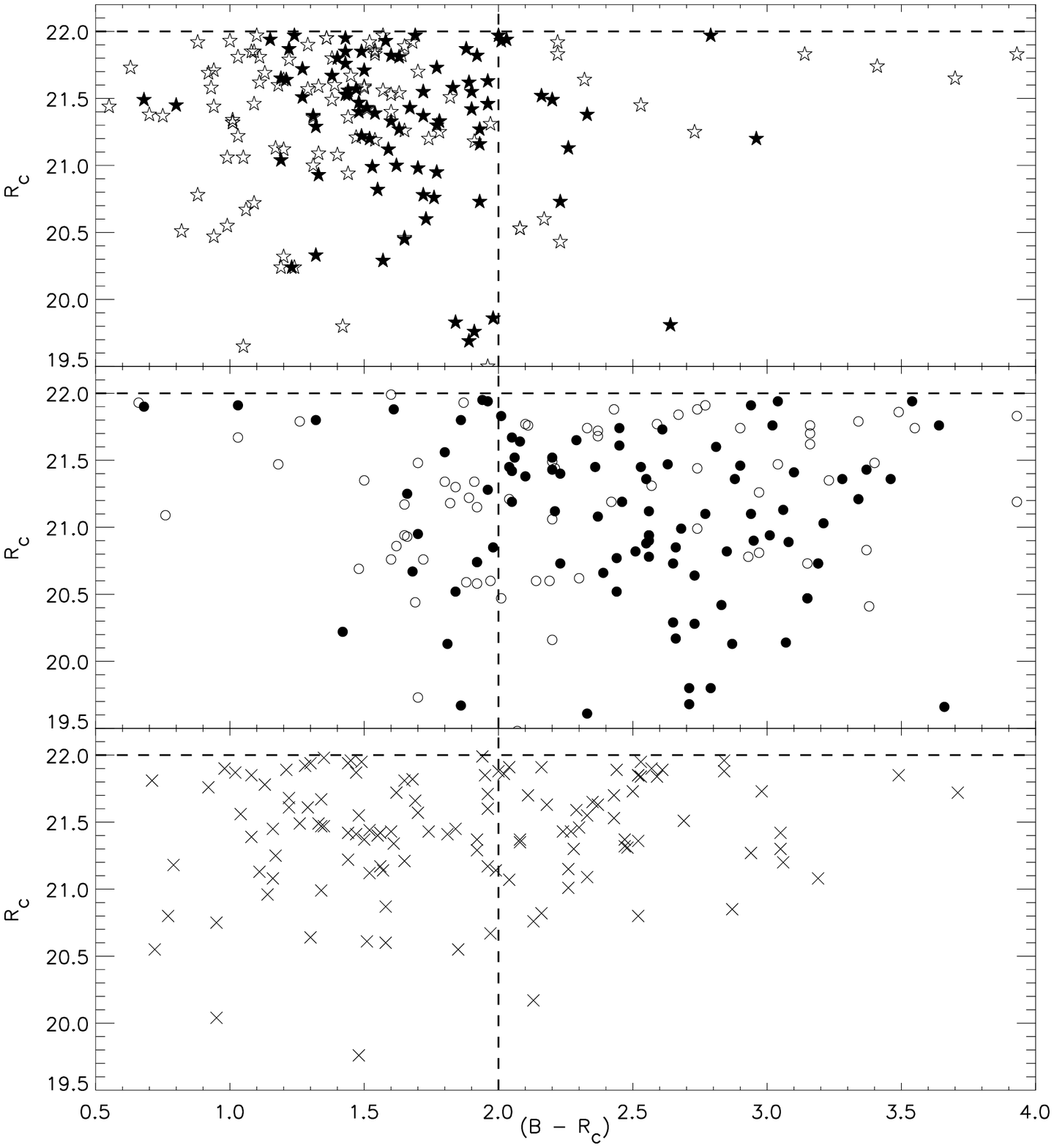,width=0.5\textwidth}}
\caption{$B-R_c$ colour vs $R_c$ magnitude for $R_C \leq 22$ objects
  targetted with Magellan. The top panel (stars) and middle panel
  (circles) show the positions of the 187 emission redshift objects and
  196 absorption redshift objects in this plane respectively. Filled
  symbols represent objects which make the redshift and luminosity cuts
  for our analysis ($\sim 50$ per cent of all objects in the range $0.3
  \leq z \leq 0.55$; \MBj\ $\geq -18.5$). The bottom panel locates in
  the colour-magnitude plane the 124 objects targetted with Magellan
  which failed to yield redshifts. The magnitude limit ($R_c = 22$) and
  a rough division in colour at $B-R_c = 2.0$ are also shown (dotted
  lines).} 
\label{fig:colmagZcompl}
\end{figure}

To investigate redshift incompleteness in the galaxies targetted with
Magellan, we examine the distribution of targetted galaxies in
colour-magnitude ($(B-R_{c}),R_{c}$) space
(Figure~\ref{fig:colmagZcompl}). If there were a bias associated with
preferentially losing absorption-line (early-type) galaxies, then one
would expect those galaxies for which we could not measure redshifts to
be clustered mainly towards the red end of the colour range. In
Figure~\ref{fig:colmagZcompl} we compare the colour and magnitude
distributions of objects for which we could not measure redshifts
(bottom panel - hereafter referred to as lost objects) with those for
which we have emission redshifts (top panel) and absorption redshifts
(centre panel). We restrict this analysis to fainter than $R_{c}=19.5$
where not all targets are successful. Open symbols represent objects
with redshifts outside our range of interest, $0.3 \leq z \leq 0.55$ or
below the luminosity range considered in the analysis, \MBj $\geq
-18.5$. Lost objects are not concentrated in the red ($B-R_c \gsim
2.0$), absorption redshift dominated region, as might be expected if we
were preferentially losing absorption line ($\sim$ early type)
galaxies. Indeed, the lost galaxies are spread quite evenly across the
full colour range, suggesting no colour bias due to
redshift-incompleteness (and thus no bias in star forming
properties). Nonetheless, we are cautious in evaluating any such bias
as it will impact directly upon our results. Therefore we perform two
further checks. 

The first check involves making a detailed analysis of the lost
objects. We begin by making a manual examination of the spectra for the
19 lost objects with magnitudes $R_c \leq 21$. Six of these objects are
stars misclassified as galaxies and are removed from the catalogue. A
further 10 of the objects have noisy, low signal spectra, which could
represent Low Surface Brightness (LSB) galaxies or astrometric offsets
from the brightest galaxy regions. Most of these objects have extended
morphology, as assessed from the CNOC2 imaging. We also characterize
the objects by measuring the signal in a significant part of the
spectrum. We measure the combined signal (total counts, \emph{cts})
from the wavelength ranges $5300$\AA$-5530$\AA\ and
$5645$\AA$-5820$\AA\ which span the most efficient region of the
grism, eliminating the strong night sky lines. From 13 lost objects
with $cts \geq 3\times10^{4}$, there are 8 stars, 1 quasar, 1 BL Lac, 1
LSB galaxy, 1 highly noisy spectrum and 1 unclassifiable object. In the
low signal range, most objects with $cts \leq 10^{4}$ are lost,
although there are also some relatively low signal to noise galaxies
with clear absorption redshifts. Finally, lost objects in the
intermediate $10^{4} \leq cts \leq 3\times10^{4}$ range tend to have
noisy spectra, often lacking emission lines. However, all blue
galaxies ($B-R_c \lsim 2.0$) with redshifts in this region possess
emission lines and so it seems possible that the blue lost objects in
this region lie at low redshift where [OII] falls out of the spectral
window. Alternatively, it is possible that those galaxies without
redshift do not for some reason possess emission lines despite their
blue colours. However, red ($B-R_c \gsim 2.0$) lost objects with this
signal range simply appear noisier and with less obvious absorption
features than their counterparts with redshifts. 

\begin{figure}
\centerline{\psfig{figure=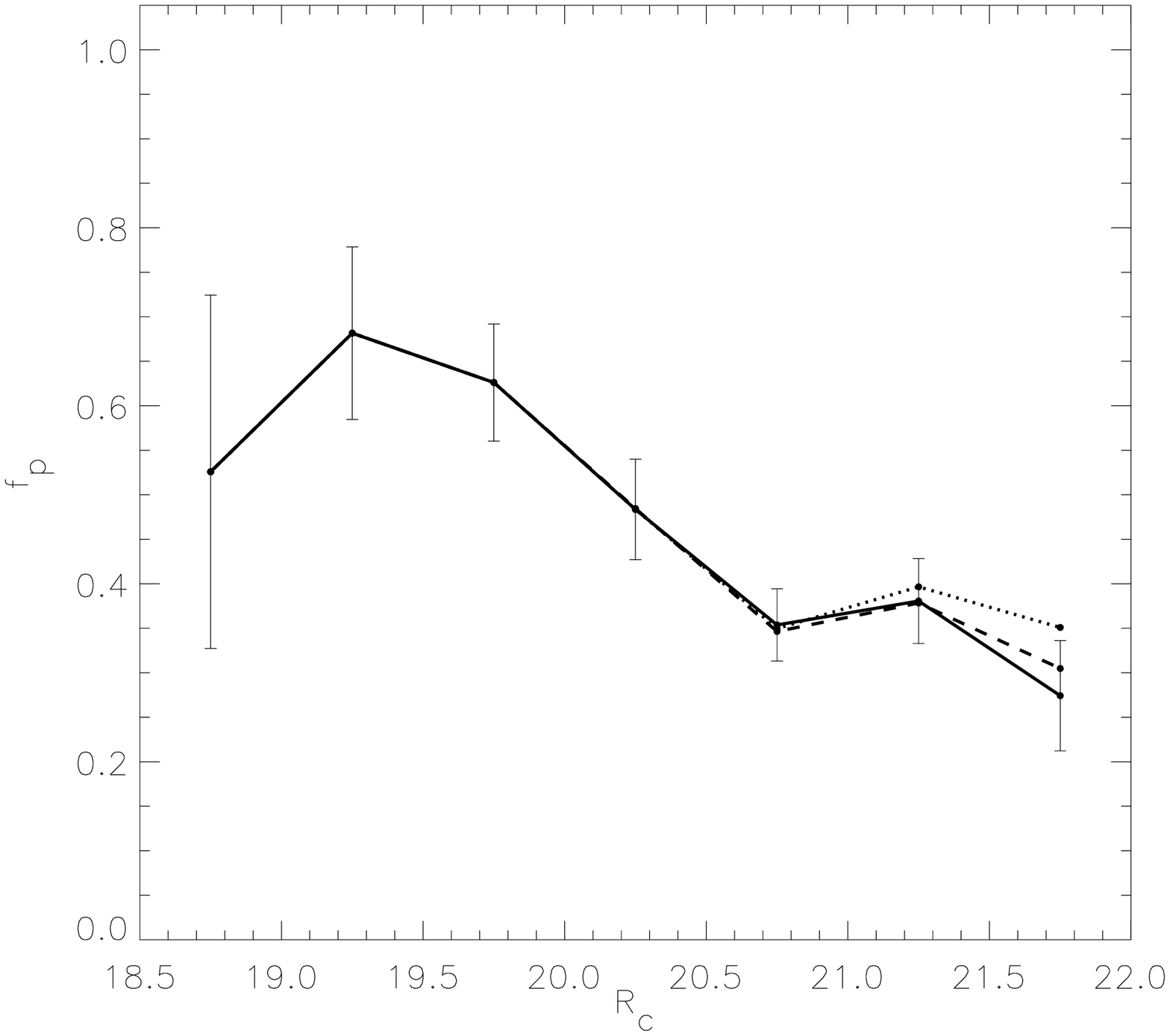,width=0.5\textwidth}}
\caption{The fraction of passive galaxies, $\fp$ as a function of
  apparent magnitude for all galaxies in the range $0.3 \leq z \leq
  0.55$. Galaxies are weighted by $W_C$ to account for selection
  bias. The solid line does not make any correction to account for bias
  in Magellan redshift incompleteness. Jackknife errors are computed in
  this case. The dashed line represents the case where galaxies with
  Magellan redshifts have also been weighted by an additional weight
  $W_{NoZ}$, effectively resampling Magellan lost galaxies (no
  redshifts) in colour-magnitude space. The dotted line case applies
  this additional weight only to red ($B-R_c \geq 2$) galaxies.} 
\label{fig:fpmagnoZ} 
\end{figure}

The second check involves computing a redshift incompleteness weight,
$W_{NoZ}$ which is computed by distributing the weight of the lost
object across its five nearest neighbours with redshifts in colour
($B-R_c$) - magnitude ($R_c$) space. Nearest neighbours are computed by
equating 1 magnitude in $R_c$ to 0.5 magnitudes in colour. Then the
combined weight of each galaxy is computed to be $W = W_{rad} \times
W_{mag} \times W_{NoZ}$. We note that \citet{Wilson02} have performed a
similar correction to compensate for galaxies in their sample for which
they could not obtain redshifts. In Figure~\ref{fig:fpmagnoZ} we show
the fraction of passive galaxies in the sample $\fp$ (at $0.3 \leq z
\leq 0.55$) as a function of apparent magnitude $R_c$ without applying
this additional weight $W_{NoZ}$ (solid line) with statistical errors
computed using the Jackknife method. We show how this uncorrected
version changes when we apply the redshift incompleteness weight
($W_{NoZ}$) to all galaxies (dashed line) and when we apply it to red
galaxies only ($B-R_c \geq 2.0$, dotted line). The latter correction
assumes that all lost blue galaxies lie outside our redshift range
whilst lost red galaxies are lost simply due to low signal to noise
spectra. This is the most biased scenario we can imagine, yet even in
the faintest bin, the difference made to $\fp$ is comparable with the
statistical error. Therefore we consider the LDSS2 redshift
incompleteness to be unbiased at $R_C \leq 22.0$, to the best of our
knowledge.  

\end{document}